\documentclass[12pt]{article}
\usepackage{geometry}[1in 1in 1in 1in]
\usepackage{graphicx} 
\usepackage{amsmath,amssymb,amsbsy,amstext,amsthm,IEEEtrantools,cite}
\usepackage[table,xcdraw,dvipsnames]{xcolor} 
\usepackage{braket}
\usepackage{hyperref}
\hypersetup{
    colorlinks=true,
    citecolor=blue,
    urlcolor=blue,
    linkcolor=blue,
}
\usepackage{float}
\usepackage{tikz, subcaption,quantikz}
\usetikzlibrary{decorations.pathreplacing}
\usepackage{setspace}
\usepackage{listings}
\usepackage{authblk}

\usepackage{grffile}[=v1]
\graphicspath{{figs}}
\newcommand{\lbow}{\hspace{-3pt} \parbox{8pt}{\tikz{\draw[<-,line width=1pt] (0,0) -- (12pt,0)}} \hspace{-8pt} \left\{ \hspace{-4pt} \left\vert}
\newcommand{\rbow}{\right\vert \hspace{-4pt} \right\} \hspace{-13pt} \parbox{10pt}{\tikz{\draw[->,line width=1pt] (0,0) -- (12pt,0)}}}

\newcommand{\ketbra}[2]{\left| #1 \right\rangle \left \langle #2 \right|}
\newcommand{\BraKet}[2]{\langle #1 \vert #2 \rangle}

\newcommand{\TR}{{\rm Tr}}
\DeclareMathOperator{\sign}{sign}

\title{Quantum Circuits for SU(3) Lattice Gauge Theory}

\author[1]{Praveen Balaji}
\author[1]{Cian\'an Conefrey-Shinozaki}
\author[1]{Patrick Draper\thanks{pdraper@illinois.edu}}
\author[1]{Jason K. Elhaderi}
\author[1]{Drishti Gupta}
\author[1]{Luis Hidalgo}
\author[1]{Andrew Lytle}
\author[2]{Enrico Rinaldi}
\affil[1]{Department of Physics, University of Illinois, Urbana, IL 61801}
\affil[2]{Quantinuum K.K., Otemachi Financial City Grand Cube 3F\\
1-9-2 Otemachi, Chiyoda-ku, Tokyo, Japan}

\begin{document}

\maketitle

\begin{abstract}
    Lattice gauge theories in varying dimensions, lattice volumes, and truncations offer a rich family of targets for Hamiltonian simulation on quantum devices. In return, formulating quantum simulations can provide new ways of thinking about the quantum structure of gauge theories. 
    In this work, we consider pure $SU(3)$ gauge theory in two and three spatial dimensions in a streamlined version of the electric basis. We use a formulation of the theory that balances locality of the Hamiltonian and size of the gauge-invariant state space, and we classically pre-compute dictionaries of plaquette operator matrix elements for use in circuit construction. We build circuits for simulating time evolution on arbitrary lattice volumes, spanning circuits suitable for NISQ era hardware to future fault-tolerant devices.  Relative to spin models, time evolution in lattice gauge theories  involves more complex local unitaries, and the Hilbert space of all quantum registers may have large unphysical subspaces.  Based on these features, we develop general, volume-scalable tools for optimizing circuit depth, including pruning and fusion algorithms for collections of large multi-controlled unitaries. We describe scalings of quantum resources needed to simulate larger circuits and some directions for future algorithmic development.
    
\end{abstract}
\newpage
\tableofcontents
\newpage
\section{Introduction}
Rapid advances in quantum computing technology have led to renewed interest in the Hamiltonian simulation of fundamental physics. The classical simulation of many quantum systems is believed to be intractable, due in part to the size of the systems' state spaces, which may require exponential classical resources to store and manipulate. Quantum computers,  utilizing native quantum degrees of freedom, offer an exciting alternative.

Gauge theories underpin the Standard Model of particle physics and have long been studied by theoretical, experimental, and computational means. The Hamiltonian formulation of lattice gauge theories was first developed in~\cite{Kogut:1974ag}. In the intervening decades, much of the computational work focused instead on the Lagrangian formulation and numerical path integral evaluation. In cases where the Euclidean action is real and bounded from below, the path integral weight can be interpreted as a probability, making these theories amenable to (classical) Monte Carlo techniques. The most well-known and widely studied example is lattice quantum chromodynamics (QCD), where high-performance classical simulations have successfully calculated hadronic masses \cite{bulava:2022:hadron_spectroscopy}, form factors \cite{meyer:2022:form_factors}, and other equilibrium properties to high precision \cite{USQCD:2022mmc,FlavourLatticeAveragingGroupFLAG:2024oxs}. Despite these advances, many properties of gauge theories have remained difficult or intractable to study classically, including general scattering processes, non-equilibrium dynamics, and physics at large baryon chemical potential or topological $\theta$ angle.  Quantum computing opens up a new avenue by which lattice gauge theories may be studied in the Hamiltonian formulation \cite{ Bauer:2023qgm,bauer:2023:quantum_simulation,di_meglio:2024:qc4hep}. In light of recent technological progress, the problem of mapping gauge theory degrees of freedom and the Hamiltonian to those of quantum hardware platforms has been of great interest~\cite{Ciavarella:2021nmj,Alam:2021uuq,Davoudi:2022xmb,bauer-2023-hamilton-su,Zhang:2023agx,Davoudi:2024wyv,Grabowska:2024emw,Lamm:2024jnl,Assi:2024pdn,Burbano:2024uvn,Kurkcuoglu:2024cfv, jakobs:2025:dynamics,ale:2024:su2lattice,halimeh:2024:universal,Fontana:2024rux}. 

A first step towards placing the Standard Model on a quantum computer is to develop real-time simulations of QCD or one of its simpler relatives, including pure $SU(2)$ or $SU(3)$ gauge theory, or QCD in lower dimensions. 
Several approaches to the quantum simulation of $SU(3)$ gauge theory have been explored; see e.g.~\cite{Byrnes:2005qx, Ciavarella:2021nmj, Kadam:2022ipf,ciavarella:2023:improved_hamiltonians, farrell:2023:preparations, atas:2023:tetra_pentaquarks,kan:2022:lattice_qed_qcd, fromm:2024:strong_coupling,Gustafson:2024kym, gustafson:2024:su3_discrete_subgroup, ciavarella:2024:large_N, Kadam:2024zkj,bergner:2024:orbifold, Than:2024zaj,kadam:2024:loop_string_hadron}. In the future, some hybridization of the techniques developed in these studies may provide the earliest exhibitions of ``quantum advantage for particle physics."

In this work, we develop circuits and tools for simulating pure $SU(3)$ gauge theory on general-purpose qubit-based quantum computers. Although the underlying technology used to implement qubits can vary (trapped ions, photonic, Rydberg atoms, superconducting transmons, etc.), the circuit-level specifications of time evolution operators that we develop here can be deployed on any of these devices.

We work in a reduced version of the electric basis, in which the basic degrees of freedom are gauge fluxes on links, supplemented by singlet information encoded on vertices. Because of the latter, it is a generalization of what is known as the local multiplet basis in the literature. The plaquette-local Hamiltonian is computed classically using the representation theory of $SU(3)$, ensuring that only physical transitions that satisfy the Gauss law constraint are present at the noiseless level. By specifying the transitions locally, with a local Hilbert space that grows with the size of the electric truncation, the overall quantum resources (qubits and gates)  scales with the size/volume of the system.
Our specification is complete, taking into account subtleties associated with different multiplicities of gauge-invariant states that can arise at each vertex.

We have developed a modular code that generates Hamiltonian evolution circuits for the $SU(3)$ gauge theory, with a specified truncation to render the Hilbert space finite-dimensional, and in 1, 2, or 3 spatial dimensions, with lattice volume and truncations specified by the user. Our circuits are based on Givens rotations that drive transitions between any two physical bitstrings. We have developed and implemented various optimizations to reduce overall circuit depths, including strategies that remove controls from rotation gates which are redundant on the physical subspace of the simulation Hilbert space. These optimizations typically reduce circuit depths by a factor of 5-10. The code is being assembled into a Python package which we plan on open-sourcing in the near future, and will include additional tools supporting the downstream analysis of simulation results such as the reconstruction of physical lattice states, the computation of simple electric-basis observables from shots data, and options for performing rudimentary error mitigation.

The remainder of this article is organized as follows. In Sec.~\ref{sec:theoretical setup} we construct the reduced electric basis,  which balances locality of the lattice Hilbert space while preserving gauge invariance. We provide a ``master formula" for the matrix elements of the plaquette operator in Eqs.~(\ref{sitefactors}),(\ref{masterform}) and describe its efficient classical pre-computation. In  Sec.~\ref{sec:quant sim} we build quantum circuits using a general approach that applies to any lattice and truncation. We describe several optimizations and provide resource estimates for a variety of small lattices in Table~\ref{gatestable}. In Sec.~\ref{results} we perform classical statevector simulations for small lattices in $d=3/2$, $d=2$, and $d=3$ ($d=3/2$ refers to a line of plaquettes) with different boundary conditions. We study effects of varying the coupling, irrep truncation, number of Trotter steps, and approximation on magnetic transitions. For the $2\times 2$ lattice in $d=2$, we also perform noisy simulations using the Quantinuum System Model H2 emulator~\cite{Quantinuum}. For the $2\times 2\times2$ lattice in $d=3$, we perform a simple tensor network analysis using matrix product states. Finally, in Sec.~\ref{future} we conclude with a summary of some directions for future study. 

\section{Theoretical setup}
\label{sec:theoretical setup}
Our target theory is given by the Kogut-Susskind (KS) Hamiltonian, describing $SU(N)$ gauge theory discretized on a rectilinear spatial lattice:
\begin{align}
  \label{eq:ks hamiltonian def}
  H = H_E + H_M = \frac{1}{2a} \biggl[ g^2 \sum_{\ell\in \mathcal{L}} E_{\ell}^2  + \frac{1}{g^2} \sum_{P\in \mathcal{P}} 2N - \Box_{P} - \Box_{P}^\dag \biggr].
\end{align}
$g$ is the gauge field coupling, and $a$ is the lattice spacing. Throughout this work, we set $a=1$. $\mathcal{L}$ denotes the set of all links and $\mathcal{P}$ the set of all plaquettes of the lattice. We work in a  version of the electric basis, in which the electric operator $E_{\ell}^2$ is diagonal. The  magnetic Hamiltonian on plaquette $P$ is defined in terms of fundamental-representation link operators $U^f$ circumscribing $P$:
\begin{equation}
  \label{eq:plaquette def}
\Box_{P} \equiv \TR(U^f(s, \vec e_i)U^f(s + \vec e_i, \vec e_j)U^f(s + \vec e_j, e_i)^{\dagger}U^f(s, \vec e_j)^{\dagger}).
\end{equation}
Here $s$ denotes a lattice site at a corner of $P$ and $\vec e_{i,j}$ are orthonormal lattice vectors. The trace is taken in the defining representation.
Of fundamental importance will be the classical pre-computation of matrix elements of $\Box_{P} + \Box_{P}^\dag$ in the gauge-invariant basis. First we describe the construction of this basis.

\subsection{Reduced electric basis states}

We begin in the  Hilbert space associated with the standard electric basis~\cite{Byrnes:2005qx}. (Subsequently we will refer to the standard basis as just the \emph{electric basis}, to be distinguished from the ``gauge-invariant electric basis" and ``reduced electric basis" defined below.) 
The whole-lattice Hilbert space is constructed as a tensor product of Hilbert spaces associated with each link. For a single link $\ell(s, \vec{e}_{i})$, the electric basis states $\ket{R, \lambda_-, \lambda_+}$ are labeled by an $SU(N)$ irrep $R$ and two multi-indices $\lambda_{+/-}$. The latter represent two states in the irrep $R$, associated with the two ends of the link $\ell$, respectively. In $SU(3)$, for example, an irrep can be labeled by whole numbers  $R = \left( p, q \right)$, and the value of the quadratic Casimir operator is given by 
\begin{equation}
  \label{eq:quadratic casimir eigenvalue}
E^2\ket{R, \lambda_-, \lambda_+} = \frac{1}{3} \left( p^2 + q^2 + pq + 3 \left( p + q \right) \right)\ket{R, \lambda_-, \lambda_+}.
\end{equation}

It is  notationally convenient to split the link state into a product of two ``half-link" states (this terminology was also used in~\cite{Burbano:2024uvn}). 
For each link $\ell(s, \vec{e}_i)$ in the lattice, we introduce the half-link states $\ket{R_{\ell}, \lambda_{\ell^{\pm}}}$. The ``positive'' and ``negative'' half-links are defined according to the unit vector $\vec{e}_i$; the positive half-link $\ell^{+}$ is the end of the link toward which $\vec{e}_i$ points, and the negative half-link $\ell^{-}$ is the end toward which $-\vec{e}_i$ points.  Electric basis states of $\ell$ are then written in terms of half-link states as
\begin{equation}
  \label{eq:full link wrt half link def}
\ket{R_{\ell}, \lambda_{\ell^{-}},\lambda_{\ell^{+}}} \equiv \ket{R_{\ell}, \lambda_{\ell^{-}}} \otimes \ket{R_{\ell}, \lambda_{\ell^{+}}}.
\end{equation}
This notation will be useful later to simplify the expression for the plaquette operator matrix elements. 

Even for a single link, the dimension of the Hilbert space is infinite.  A finite-dimensional Hilbert space is obtained by truncating the electric basis to irreps of low-lying electric energy density $\propto E^2$.  We denote the truncation to $r$-index tensor irreps by $T_{r}$. For instance,  $T_1$ corresponds  to the truncation to irreps $\{1,3,\bar 3\}$ in $SU(3)$.

This construction, detailed in~\cite{Byrnes:2005qx}, has the significant advantage that it expresses the lattice Hilbert space as a tensor product of local Hilbert spaces. Since the Hamiltonian is also local at the level of plaquettes, the construction has good ``in-principle" scalability properties with the lattice volume. Its primary disadvantage is that the dimension of the local Hilbert space grows rapidly with irrep truncation, and in a way not particularly naturally adapted to qubits, such that the plaquette operator becomes rather complicated to implement. Furthermore most of the lattice Hilbert space is unphysical, since the Gauss' law has not been imposed. 

A spatial gauge transformation at lattice site $s$ is generated by 
\begin{equation}
    G^a(s) = \sum_{i=1}^{3} \left[ E_+^a(s - e_i, e_i) - E_-^a(s, e_i) \right].
    \label{gauss_op}
\end{equation}
Here $E_+^a$ and $E_-^a$ are the electric field operators generating the left and right action of the gauge group on the link operators, and the $\lambda$'s in Eq.~(\ref{eq:quadratic casimir eigenvalue}),(\ref{eq:full link wrt half link def}) are the eigenvalues of the $E^a_{+,-}$ in the Cartan subalgebra. Gauss' law can then be enforced by making physical states satisfy $G^a(s) \ket{\text{phys}} = 0$,
and it is equivalent to the requirement that the half-link states of $\ket{\text{phys}}$ meeting at site $s$ form an $SU(N)$ singlet. Thus, a basis of \emph{physical} lattice states may be obtained from the electric basis by contracting the link states meeting at each vertex $s$ with an $SU(N)$-invariant tensor. Explicitly, a single physical electric basis state $\ket{\omega}$ may be expanded in terms of electric basis states\footnote{Under a gauge transformation $\Omega$, the positive and negative half-link are transformed as $\ket{R_\ell, \lambda_{\ell^+}} \to D^f_{\lambda\rho}(\Omega) \ket{R_\ell, \rho_{\ell^+}}$ and $\ket{R_\ell, \lambda_{\ell^-}} \to D^f_{\rho\lambda}(\Omega^\dag) \ket{R_\ell, \rho_{\ell^-}}$ for unitary representation matrices $D^f(\Omega)$ (see, e.g., \cite{bauer-2023-hamilton-su, zohar-2015-formulat}). The relation between the matrix elements $D^f_{\rho\lambda}(\Omega^\dag)$ and $D^{\bar{f}}_{\tilde{\lambda}\tilde{\rho}}(\Omega)$  depends on sign conventions, generalizing the Condon-Shortley phases. In general
$D^f_{\rho\lambda}(\Omega^\dag) = D^{f^*}_{\lambda\rho}(\Omega) = \phi(\lambda) \phi(\rho) D^{\bar{f}}_{\tilde{\lambda}\tilde{\rho}}(\Omega)$, where $\bar{f}$ and $\tilde{\lambda}$ are defined in the text above and the $\phi$'s are the conventional phases. We discuss them more thoroughly around Eqs.~(\ref{signseq1}) and (\ref{signseq2}). This distinction between the basis vectors of barred irreps and naive conjugation is the origin of the phases in Eq.~(\ref{gauge_invariant_state}).} using generalized Clebsch-Gordan coefficients (CGCs):
\begin{equation}
    \ket{\omega} = \bigotimes_{s \in \mathcal{S}} \sum_{\lambda} \left( \textstyle\prod_{\ell^-(s)} \phi(\lambda) \right) \BraKet{1,\Gamma_s}{ \otimes_{\ell^-(s)} (\bar{R}, \tilde{\lambda}) \otimes_{\ell^+(s)} (R, \lambda)} \ \ket{\otimes_{\ell(s)} (R,\lambda)}.
    \label{gauge_invariant_state}
\end{equation}
 Here $\ell(s)$ denotes the set of half-links meeting at site $s$, and $\ell^\pm(s)$ are the subsets of positive/negative half-links at the site.   $\bar{R}$ is the  representation conjugate to $R$; if $R$ is a real representation, then $\bar{R}$ is the same as $R$. Apart from possible phase conventions $\phi(\lambda)$ which will be revisited below (see footnote), $\tilde{\lambda}$ is the basis state of $\bar{R}$ obtained by negating the weights of the state $\lambda$ of $R$.  
 The CGC $\BraKet{1,\Gamma_s}{ \otimes_{\ell^-(s)} (\bar{R}, \tilde{\lambda}) \otimes_{\ell^+(s)} (R, \lambda)}$ is the $\Gamma^{\rm th}$ invariant tensor in the corresponding product of irreps. 
The allowed invariant tensors at each vertex depend only on  the link irreps meeting at the vertex, but in general, they are not unique; for example, in $SU(3)$ there are two singlets in $8\otimes 8\otimes 8$ and 145 in $8\otimes 8\otimes 8\otimes 8\otimes 8\otimes 8$.  We discuss the enumeration of singlets further in section \Ref{cgsec}. 
 
Thus, a complete specification of a gauge invariant basis state is given by an assignment of irrep to every link along with a consistent choice of singlet at every vertex. In particular, we no longer need separate basis states for each irrep internal state (e.g. red, green, blue for $R=3$.) The \emph{gauge-invariant electric basis} spans the physical Hilbert space, which is much smaller than the Hilbert space spanned by the electric basis.

The fact that the consistent vertex singlets are dependent on the link irreps is inconvenient for locality and for implementing the magnetic time evolution operator. Consistent assignments of irreps to every link and singlets to every vertex is essentially a global problem.  For this reason, it is advantageous to relax the
``consistency" requirement, and simply define a \emph{reduced electric basis} of states by an assignment of irrep to every link and a choice of ``singlet multiplicity index" at every vertex. In this construction, the Hilbert space is built again from a tensor product of local Hilbert spaces. However, basis states are included regardless of whether the tensor product of the link irreps contains even one singlet, or of whether the  singlet multiplicity index is in the range of the number of singlets in the tensor product. Thus some unphysical states are reintroduced.

The idea of using link irreps to define the Hilbert space, rather than the full electric basis of~\cite{Byrnes:2005qx}, was developed  for quantum simulations of gauge theories in~\cite{Banuls:2017ena,Klco:2019evd,Ciavarella:2021nmj}, and is often referred to as the ``local multiplet basis" in the literature. The reduced electric basis defined above extends the local multiplet basis to include site singlet multiplicity information, which is generically essential to obtain a complete description of the gauge invariant Hilbert space and obtain meaningful matrix elements. Some previous works, for example~\cite{Ciavarella:2021nmj}, do not have to assign separate singlet multiplicity degrees of freedom because the truncation is sufficiently strong that no multiplicity arises. In $SU(3)$, with irreps truncated to $\{1, 3, \bar 3\}$ and working with a line of plaquettes, the singlets associated with any collection of three irreps is unique. This can be extended to higher dimensions by working with lattices with only triagonal vertices (see e.g.~\cite{Kavaki:2024ijd}), but it still depends on the irrep truncation: once two-index tensor irreps are included, the singlets are no longer uniquely identified by the link irreps.     

The nice locality properties of the reduced electric basis is manifest. Let us illustrate the price --- some unphysical basis states --- with some examples. In $SU(3)$ in three spatial dimensions with link irreps drawn from $\{1,3,\bar 3\}$, the largest number of distinct invariant tensors (6) is found in $3\otimes 3\otimes 3\otimes \bar 3\otimes \bar 3\otimes \bar 3$, so the singlet multiplicity index must run from one to six. However, other link states may contain fewer singlets. For example, $3\otimes 3\otimes 3\otimes 1\otimes 1\otimes 1$ contains only one singlet, so when the links are in these irreps there are five choices of singlet multiplicity index at the vertex that are inconsistent with the links. Furthermore $3\otimes 3\otimes 3\otimes 3\otimes \bar 3\otimes \bar 3$ contains zero singlets, so the corresponding link irreps are inconsistent by themselves. Thus the reduced electric basis expands the Hilbert space to include unphysical states, but the Hilbert space dimension is still considerably lower than in the full electric basis, and the Hamiltonian is local.\footnote{It is not clear what impact these unphysical states will have on fidelity in
simulations on NISQ era devices, apart from making it worse. This must be tested on small
systems, as below in Fig~\ref{fig:noisy}, in order to develop mitigation strategies. In the fault-tolerant era, the hope
(shared for essentially any scalable lattice gauge theory simulation) is that error correction will be sufficient to prevent
wandering off into unphysical sectors.}

To summarize, a basis state $|\omega\rangle$ of the reduced electric basis is given by
\begin{align}
|\omega\rangle =\bigotimes_{s \in \mathcal{S},\ell\in\mathcal{L}} |R_\ell\rangle|\Gamma_s\rangle
\end{align}
where:
\begin{itemize}
    \item  the link irreps $R_{\ell}$ are drawn independently from the irrep truncation, and
    \item the site singlet multiplicity index $\Gamma_s$ runs from one to the maximum number of independent singlets that appear in any  tensor product of $2d$ elements of the irrep truncation.  (We may also impose an additional truncation on the collection of singlets; see Sec.~\ref{truncsec}.)
\end{itemize}
Note that the reduced electric basis includes all the basis states of the gauge-invariant electric basis exactly once.

\subsection{Matrix elements of the  plaquette operator}

We seek to classically pre-compute matrix elements of the gauge-invariant plaquette operator between gauge-invariant basis states.  This involves the action of the operator both on the link irreps and the site singlets. We begin by discussing the two types of Clebsch-Gordan coefficients that appear in the final master formula for these matrix elements.

The first type of CGC is associated with the site singlets. In $d$ spatial dimensions, at most $2d$ half-links meet at each site---in $d=3$, for example, three negative half-links and three positive half-links meet.\footnote{There will be fewer half-links meeting at sites on the boundary of nonperiodic lattices.} Let $\ell(s) = \ell^-(s)\cup \ell^+(s)$ denote the set of all negative and positive half-links meeting at site $s$, respectively. Fixing the irreps $R_{\ell(s)}$, we define a tensor product space 
\begin{align}
\otimes_{\ell\in\ell^+(s)} R_{\ell}\otimes_{\ell\in\ell^-(s)} \bar R_{\ell}
\end{align} 
 associated with half-links meeting at $s$. It is important that we impose an arbitrary but fixed ordering of the half-links meeting at each site, which we call an \textit{F-order}.  This ordering is implicit in the notation $\otimes_{\ell^\pm(s)}$. It is necessary to avoid permutation ambiguities  when the same irrep appears on multiple links meeting at a site: in general, the states appearing in an irrep direct sum decomposition of a tensor product transform nontrivially under permutations.
 
\begin{figure}[ht]
    \centering
    \includegraphics[width=\linewidth]{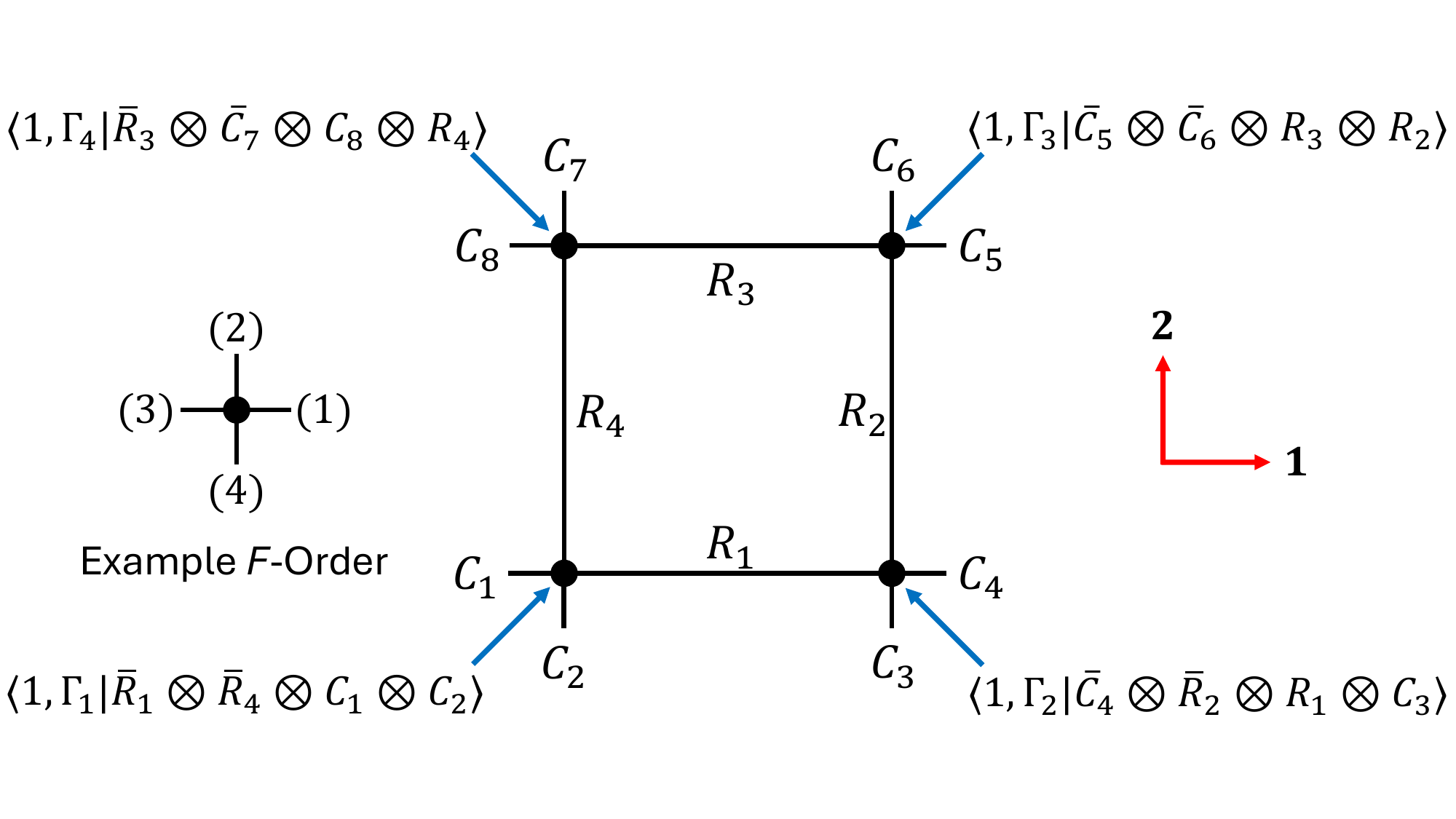}
    \caption{An $F$-ordering is a link ordering at each vertex which is used to uniquely specify invariant tensors associated with singlet states. At each vertex, the precise order in which the link irreps appear in the ``tensor product side" of the CGC is given by the $F$-ordering.}
    \label{f_order_example}
\end{figure}
 
 A standard electric basis for this space is
\begin{equation}
\bigotimes_{\ell\in\ell^+(s)} \ket{R_{\ell},\lambda_\ell}\bigotimes_{\ell\in\ell^-(s)} \ket{\bar R_{\ell},\tilde \lambda_\ell} .
\label{siteHS}
\end{equation}
Gauge-invariant states correspond to singlets in the Hilbert space spanned by~(\ref{siteHS}). The  CGCs for these singlets  were introduced already in~\eqref{gauge_invariant_state} and are denoted
\begin{align}
\braket{1,\Gamma_s|\otimes_{\ell^{-}}\bar{R}_{\ell^{-}}\tilde{\lambda}_{\ell^{-}}\otimes_{\ell^{+}}R_{\ell^{+}}\lambda_{\ell^{+}}}
\end{align}
where $\Gamma_s$ labels the $\Gamma^{\rm th}$ singlet  living in the space~(\ref{siteHS}). The half-link states appear in the tensor product in a chosen $F$-order, which means  that each $(R,\lambda)$ state appearing in a CGC can be attributed to a specific half-link on the lattice. (See Fig.~\ref{f_order_example}.)

The second type of CGC is associated with the link irreps. A  link operator in representation $r$ acts on the electric basis states of the link as
\begin{align}
  \hspace{-0.5in} &U_{\sigma\sigma'}^r \ket{R_\ell,\lambda_{\ell^-},\lambda_{\ell^+}} =\nonumber\\
&   \sum_{\Omega_\ell,\Lambda_{\ell^-},\Lambda_{\ell^+}} \sqrt{\frac{\dim(R_\ell)}{\dim(\Omega_\ell)}} \BraKet{\Omega_\ell,\Lambda_{\ell^-}}{R_\ell,\lambda_{\ell^-} ; r,\sigma} \ \BraKet{\Omega_\ell,\Lambda_{\ell^+}}{R_\ell,\lambda_{\ell^+}  ;r,\sigma'} \ \ket{\Omega_\ell,\Lambda_{\ell^-},\Lambda_{\ell^+}}.
\end{align}
In this expression, objects like $\BraKet{\Omega_\ell,\Lambda_{\ell^-}}{R_\ell,\lambda_{\ell^-} ; r,\sigma} $ denote CGCs.

The gauge invariant plaquette operator appearing in the KS Hamiltonian was defined in Eq.~(\ref{eq:plaquette def}) in terms of the fundamental representation link operators. It is convenient to convert $(U^f)^\dagger$ into $U^{\bar f}$. As mentioned previously, this relation can involve conventional signs:
\begin{align}
[(U^f)^\dag]_{\sigma\sigma'} = (U^f_{\sigma'\sigma})^* = \phi(\sigma)\phi(\sigma') U^{\bar{f}}_{\tilde{\sigma}'\tilde{\sigma}}.
\label{signseq1}
\end{align}
Here, $\phi(\sigma)=\pm 1$ relates the basis states associated with na\"ively conjugating $f$ to the actual basis states we use for $\bar f$. This map involves a (conventional) reversal of orientation of some basis vectors. Since the invariant tensor on $R\otimes  R^*$ is just the identity matrix, the $\phi$ are related to the singlet in $R\otimes \bar R$ by
\begin{align}
 \phi(\lambda)  =  \phi(\tilde{\lambda}) = \sign(\BraKet{1}{R, \lambda; \bar{R}, \tilde{\lambda}}) = \sign(\BraKet{1}{\bar{R}, \tilde{\lambda}; R, \lambda}).
 \label{signseq2}
 \end{align}
 Note that $\phi(\lambda)\phi(\lambda) = 1$.

We are now ready to construct our master formula. Label the sites and links around a plaquette $P$ counterclockwise as $s_1,\dots,s_4$ and $\ell_1,\dots,\ell_4$. The set of plaquette sites and links are $\mathcal{S}(P)$ and $\mathcal{L}(P)$. At each site, there are at most four links that are not part of the plaquette, but they must be known to calculate singlet CGCs. Following~\cite{Klco:2019evd,Ciavarella:2021nmj} we will refer to these ``control" or ``external" links.  The collection of control links at site $s_i$ is $\vec{C}_{s_i}$. (See Fig.~\ref{f_order_example}.) Let $\bra{\omega_{n,m}}$ denote two gauge-invariant electric basis states. (As discussed above, these are also a subset of the reduced electric basis states; we require matrix elements only for the former.) $\bra{\omega_n} \Box \ket{\omega_m}$ can then be found by acting on $\ket{\omega_m}$ with the four link operators, then contracting with $\bra{\omega_n}$, collapsing any sums over half-link states that are not modified (any half-links not part of the plaquette). 

Carrying out this procedure we arrive at the master formula for the plaquette operator matrix elements in the reduced electric basis. The expression is somewhat lengthy, so to preserve readability of the text we have placed it in the Appendix, Eq.~(\ref{masterformulafull}). Here we provide a compact version, as follows. 
 First, we can take a sign convention for the CGCs such that 
\begin{align}
 \BraKet{(A,a) \otimes (f,\sigma)}{B,b} = \phi(a)\phi(\sigma)\phi(b) \BraKet{(\bar{A}, \tilde{a}) \otimes (\bar{f}, \tilde{\sigma})}{\bar{B}, \tilde{b}} 
 \label{phaserelation}
\end{align}
 holds. (A similar relation from interchanging $f$ and $\bar{f}$ would also hold.) Second, we define the site factors 
\begin{IEEEeqnarray}{rCl}
\lbow \begin{matrix} R & G & S \\ f & \vec{C} & \bar{f} \\ R' & G' & S' \end{matrix} \rbow \ & = & \sum_{\sigma=1}^{\dim(f)} \sum_{r=1}^{\dim(R)} \sum_{r'=1}^{\dim(R')} \sum_{s=1}^{\dim(S)} \sum_{s'=1}^{\dim(S')} \sum_{\vec{c} \in \vec{C}} \phi(\sigma) \nonumber\\
& \times & \BraKet{R',r'}{(R,r) \otimes (f,\sigma)} \ \BraKet{S',s'}{(S,s) \otimes (\bar{f},\tilde{\sigma})} \nonumber\\
& \times & \BraKet{1,G}{(R,r) \otimes (S,s) \otimes (\vec{C},\vec{c})}_F \nonumber\\
& \times & \BraKet{1,G'}{(R',r') \otimes (S',s') \otimes (\vec{C},\vec{c})}_F.
\label{sitefactors}
\end{IEEEeqnarray}
The subscript $F$ on each singlet CGC reminds us that the irreps in the set $\{ R,S,\vec{C} \}$ should be placed into the lattice $F$-order, as described above.

\begin{figure}[ht]
    \centering
    \includegraphics[width=1\linewidth]{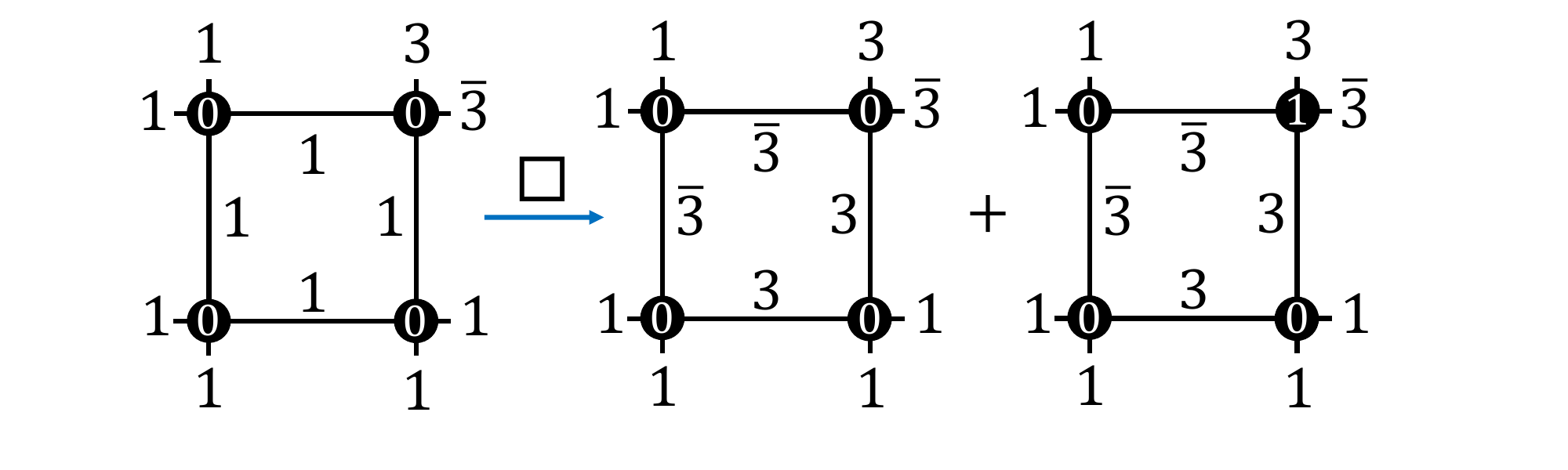}
    \caption{A representation of the action of $\square$ on a particular plaquette. Observe that initially all the vertex singlets are $0$ or unoccupied because the irreps at each vertex only decompose into one singlet irrep. However $3\otimes3\otimes\bar3\otimes\bar3$ contains two singlets in its decomposition so the operation of $\square$ will yield two plaquette states, one with nontrivial multiplicity at the top-right vertex. The amplitude for the transition to either state is given by the master formula and will in general be distinct.}
    \label{box_transition}
\end{figure}

These site factors are somewhat similar to the 9-R symbols that appear in the matrix element formulas derived in \cite{Ciavarella:2021nmj}. However, here their form is fixed for each site ($f$ always appears on the left column and $\bar{f}$ always appears on the right column), and more significantly, singlet multiplicity indices are not summed over. See Fig.~\ref{box_transition} for a sketch of how different singlet multiplicities appear in $\Box$ matrix elements. These site factors also apply to rectilinear lattices in general spatial dimensions; the only dimension dependence is in the number of irreps appearing in $\vec C$. 

Using~(\ref{phaserelation}) and~(\ref{sitefactors}), and suppressing the obvious Kronecker deltas, the master formula~(\ref{masterformulafull}) can be written in the compact form
\begin{align}
 \hspace{-0.6in} \bra{\omega_n} \Box \ket{\omega_m} =
  \left( \prod_{k=1}^{4} \sqrt{\frac{\dim(R^m_{\ell_k})}{\dim(R^n_{\ell_k})}} \right)\, &\lbow \begin{matrix} \bar{R}_{\ell_4}^m & \Gamma_{s_1}^m & \bar{R}_{\ell_1}^m \\ f & \vec{C}_{s_1} & \bar{f} \\ \bar{R}_{\ell_4}^n & \Gamma_{s_1}^n & \bar{R}_{\ell_1}^n \end{matrix} \rbow \quad \lbow \begin{matrix} R_{\ell_1}^m & \Gamma_{s_2}^m & \bar{R}_{\ell_2}^m \\ f & \vec{C}_{s_2} & \bar{f} \\ R_{\ell_1}^n & \Gamma_{s_2}^n & \bar{R}_{\ell_2}^n \end{matrix} \rbow\nonumber\\
 \times \ & \lbow \begin{matrix} R_{\ell_2}^m & \Gamma_{s_3}^m & R_{\ell_3}^m \\ f & \vec{C}_{s_3} & \bar{f} \\ R_{\ell_2}^n & \Gamma_{s_3}^n & R_{\ell_3}^n \end{matrix} \rbow \quad \lbow \begin{matrix} \bar{R}_{\ell_3}^m & \Gamma_{s_4}^m & R_{\ell_4}^m \\ f & \vec{C}_{s_4} & \bar{f} \\ \bar{R}_{\ell_3}^n & \Gamma_{s_4}^n & R_{\ell_4}^n \end{matrix} \rbow .
 \label{masterform}
 \end{align}

The factorization of these matrix elements into site factors is enormously useful for classical pre-computation of the matrix elements. Classical pre-computation begins by calculating all non-vanishing site factors, enumerating all different ways that Eq.~(\ref{sitefactors}) can be populated with irreps and singlets given a truncation. Because of the $F$-ordering imposed on the singlet CGCs, site factors corresponding to the same site but to plaquettes lying in different planes of the lattice need not be equal. Furthermore, if open boundary conditions are used, then there may be different site factors at different ``ends'' of the lattice. 

From a list of nonzero site factors, four at a time can then be ``glued'' together by making sure the columns ``match.'' More precisely, note that where the site factors abut each other in Eq.~(\ref{masterform}), the adjacent columns refer to the same active link. However, one column holds the active link irreps, while the adjacent  column in the neighboring site factor holds their conjugates. Therefore, two site factors can initially be glued together if the $S$ and $S'$ irreps of the first site factor, in the notation of Eq.~(\ref{sitefactors}), are conjugate to the $R$ and $R'$ irreps of the second. Then the rest of the site factors can be glued in this manner, forming the entire plaquette matrix element. Extra care should be taken when periodic boundary conditions constrain certain control links of separate sites to be in the same state. In a $2 \times 2$ lattice, for instance, $\vec{C}_{s_1}$ and $\vec{C}_{s_2}$ will have one link in common. This process will be repeated for plaquettes lying on different planes of the lattice (and ``ends'' of the lattice, if open boundary conditions are used.) 

Apart from the plaquette matrix elements, it is useful to generate sets of \textit{physical plaquette states}---an assignment of irrep to each of the active and control links of a single plaquette---such that each of the four sites supports a singlet, and valid assignment of singlet multiplicity index to the sites. This can be done sequentially as follows. For $s_1$, generate all possible singlets allowed by the irrep truncation, assigning  irreps to specific links.  For $s_2$ ($s_3$), connect all singlets present on $s_1$ ($s_2$) with another singlet that has an $\ell_1$ ($\ell_2$) irrep conjugate to the $\ell_1$ ($\ell_2$) irrep on $s_1$ ($s_2$). Then for $s_4$, a similar connection must be done taking into account the $\ell_3$ irrep of $s_3$ and the $\ell_4$ irrep of $s_1$. Throughout this process, boundary conditions may constrain the control link irreps present on each singlet.

\begin{table}[hb]
    \centering
    \begin{tabular}{|c|c|c|}
        \hline
        $d$ & $T_1$ & $T_2$ \\
        \hline
        $3/2$ & 81 & 28858 \\
        \hline
        2 & 19329 & $\sim 1.48 \times 10^{10}$ \\
        \hline
        3 & $\sim 2.07 \times 10^{10}$ & $\gtrsim 10^{20}$ \\
        \hline
    \end{tabular}
    \caption{The general number of $\Box$ matrix elements in different lattices and truncations with periodic boundary conditions in $SU(3)$. $T_1$ is the irrep truncation $\{ 1,3,\bar{3} \}$, and $T_2$ is the irrep truncation $\{ 1,3,\bar{3},6,\bar{6},8 \}$.}
    \label{tblphyscounts}
\end{table}

We see in Table~\ref{tblphyscounts} a rapid growth in the number of $\Box$ matrix elements in $SU(3)$ with dimension and irrep truncation. Fortunately, the plaquette operator matrix elements are very sparse, and we will provide some evidence that only a small subset of these transitions are needed to accurately capture dynamics.

 \subsection{Numerical computation of Clebsch-Gordan coefficients}
\label{cgsec}

The site factors~(\ref{sitefactors}) appearing in the master formula~(\ref{masterform}) for the physical state matrix elements of the magnetic Hamiltonian are composed of sums of $SU(N)$ Clebsch-Gordan coefficients, for example, quantities of the form 
\begin{align}
  C^{M, \alpha}_{M_1, M_2,\dots } = \braket{M \!, \alpha| M_1 \otimes M_2\otimes\dots} \label{eq:tensor product orthonormality def}
\end{align}
where each of $M, M_1,M_2,\dots $  denotes  both an irrep and a state within that irrep, and $\alpha$ indexes the outer multiplicity of $M$ in the direct sum decomposition. To find the CGCs, we use the  method described in \cite{Alex:2010wi} for products of two irreps and generalize it recursively to products  of arbitrary numbers of irreps. We briefly describe the algorithm here, and refer to \cite{Alex:2010wi} for details.

\begin{enumerate}
    \item The tensor product of irreps from the input are decomposed into a direct sum of irreps using the Little-Richardson rule, which involves constructing and composing the Gelfand-Tsetlin (GT) patterns of  the tensored irreps to obtain new GT patterns associated with irreps in the direct sum decomposition. 
   \cite{Alex:2010wi} offers an iterative algorithm based on this rule that is utilized in our code. 

    \item For each irrep in the decomposition, the CGCs of the highest weight state are found. Simultaneous eigenstates of $N-1$ commuting matrices ($J^{(l)}_{z}, 1 \! \leq \! l \! \leq \! N-1$) can be raised or lowered by generalized ladder operators,  $J^{(l)}_{+}$ and $J^{(l)}_{-}$. The highest weight state $H$  is defined as the state $\ket{H, \alpha}$ such that
    \begin{align}
      J^{(l)}_{+}\ket{H, \alpha} =  \sum_{M_1,M_2}C^{H, \alpha}_{M_1,M_2}\left(J^{S_1}_{+} \otimes \mathbb{I}^{S_2} + \mathbb{I}^{S_1} \otimes J^{S_2}_{+} \right)\ket{M \otimes M'} = 0
      \label{highest_weight_condition}
    \end{align}
     Here $S_{1,2}$ denote the irreps to which the states $M_{1,2}$ belong. Since $\left\{\ket{M_1 \otimes M_2}\right\}$ forms an orthonormal basis, Eq.~(\ref{highest_weight_condition}) gives a solvable system of linear equations for the CGCs. The number of independent solutions for the highest weight CGCs for a given irrep corresponds to its outer multiplicity in the decomposition. Since the quantities we calculate differentiate between copies of the same irrep, we construct an orthonormal basis of solutions.

    \item Using the above $J^{(l)}_{-}$ lowering operators, the CGCs for the lower-weight states are constructed from the known CGCs of the higher-weight states. With the selection rules of the lowering and raising operators, states of the same weight have the same  ``parent" states that give rise to superpositions of those states once lowered. It then becomes a matter of solving linear sets of equations for  their CGCs, as in the second step.
\end{enumerate}

We have implemented this algorithm as a supplementary Python package to our primary simulation codebase and plan to make it open-source in the future.

\subsection{Truncations}
\label{truncsec}

Thus far the formalism described applies to general $SU(N)$ gauge group, apart from explicit examples given for $N=3$. Going forward we restrict attention to $SU(3)$.

We employ several types of truncations in order to map the full lattice $SU(3)$ theory to approximations of it that are more tractable for present quantum hardware. For notational convenience we define $T_i$ to refer to the $i$-index irrep truncation; e.g., $T_1=\{1,3,\bar 3\}$ (one-index tensors),  $T_2=\{1,3,\bar 3,6,\bar 6, 8\}$ (two-index tensors), etc. These truncations can be viewed as bounding the  electric energy density on a link, cf.~(\ref{eq:quadratic casimir eigenvalue}).

In the $T_i$, no truncation of the site singlets is assumed beyond those imposed automatically by the link irrep truncation. However, we will also make use of a further truncation $T_1'$, where the irreps are truncated at $\{1,3,\bar 3\}$ and the site singlets are limited so that only two link irreps entering the site are allowed to be nontrivial at a time. This truncation can also be viewed as cutting off the electric energy density at a site.

These approximations to the full theory can be systematically improved in the obvious way. Their main value is to limit the size of the local Hilbert space associated with a plaquette, and thus, in turn, to limit the complexity of magnetic time evolution. However, these truncations may also fail to lower gate counts to a sufficient degree on present hardware. In this case it is useful to impose a further cutoff on the individual transitions induced by the magnetic Hamiltonian.  We discuss a truncation scheme for the magnetic matrix elements further in Sec.~\ref{results}.

\begin{table}[hb]
    \centering
    \begin{tabular}{|c|c|c|}
        \hline
        $d$ & $T_1$ & $T_2$ \\
        \hline
        $3/2$ & 4 & 16 \\
        \hline
        2 & 6 & 66 \\
        \hline
        3 & 28 & 1646 \\
        \hline
    \end{tabular}
    \caption{The number of unique singlets found at a site.}
\end{table}

\subsection{Magnetic matrix element distributions}

\begin{figure}[ht]
    \centering
    \begin{subfigure}[t]{0.5\linewidth}
        \centering
        \includegraphics[width=\linewidth]{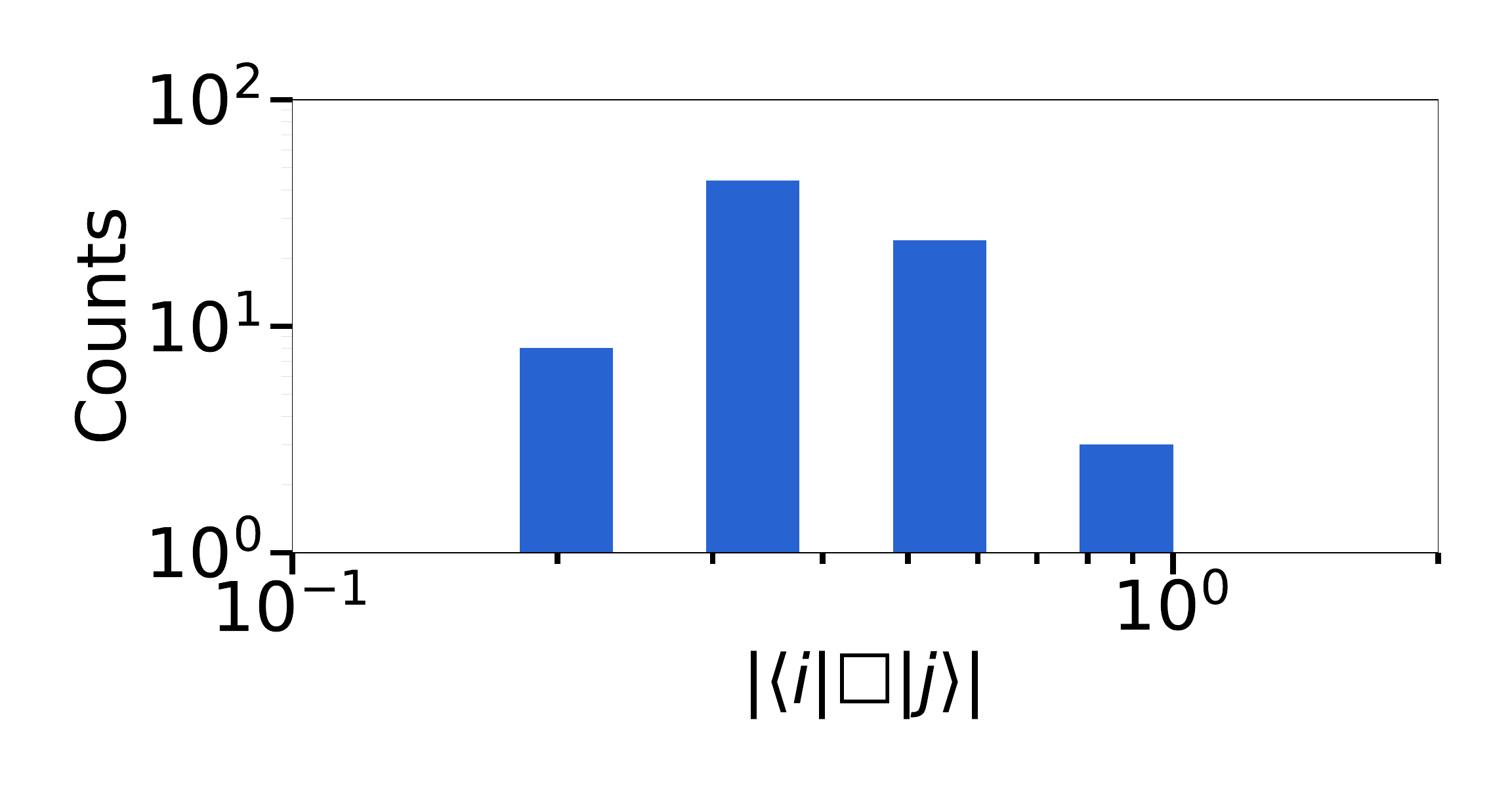}
    \end{subfigure}%
    ~
    \begin{subfigure}[t]{0.5\linewidth}
        \centering
        \includegraphics[width=\linewidth]{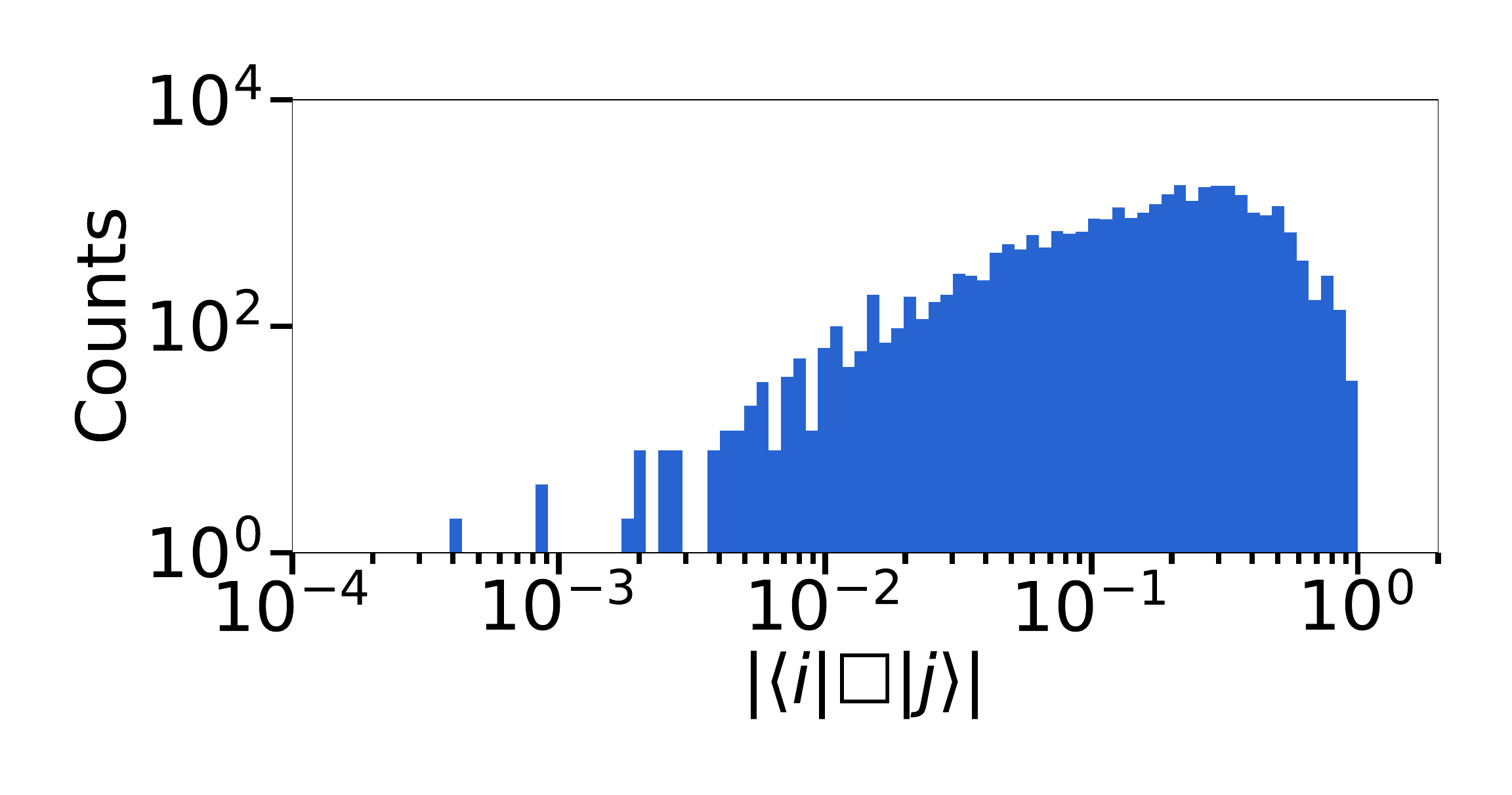}
    \end{subfigure}%
    \caption{Distributions of magnitudes of matrix elements of the plaquette operator  for the ``line of plaquettes" lattice, with $T_1$ truncation (left) and $T_2$ truncation (right).}
    \label{figd32distr}
\end{figure}

\begin{figure}[ht]
    \centering
    \begin{subfigure}[t]{0.5\linewidth}
        \centering
        \includegraphics[width=\linewidth]{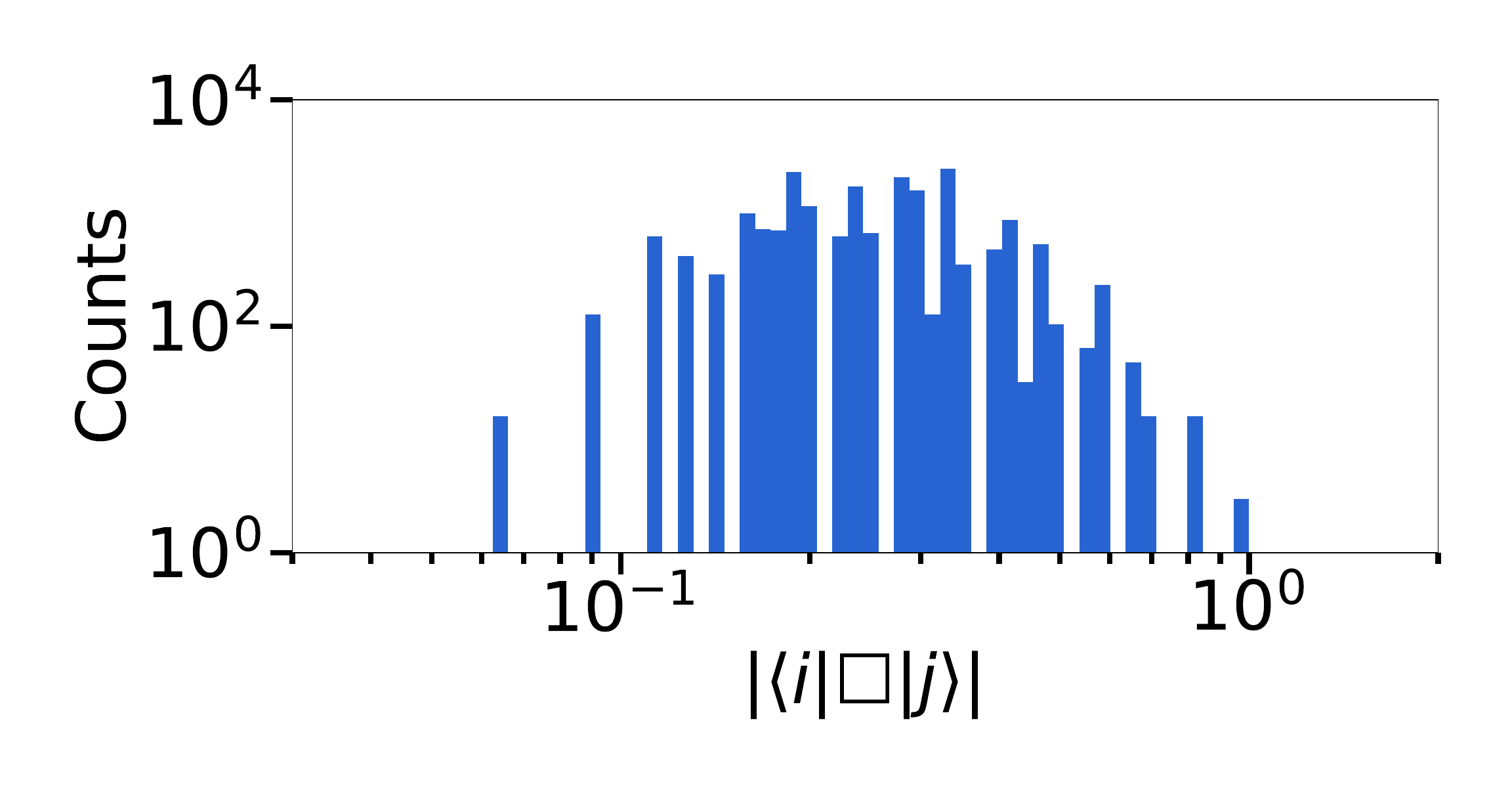}
    \end{subfigure}%
    ~
    \begin{subfigure}[t]{0.5\linewidth}
        \centering
        \includegraphics[width=\linewidth]{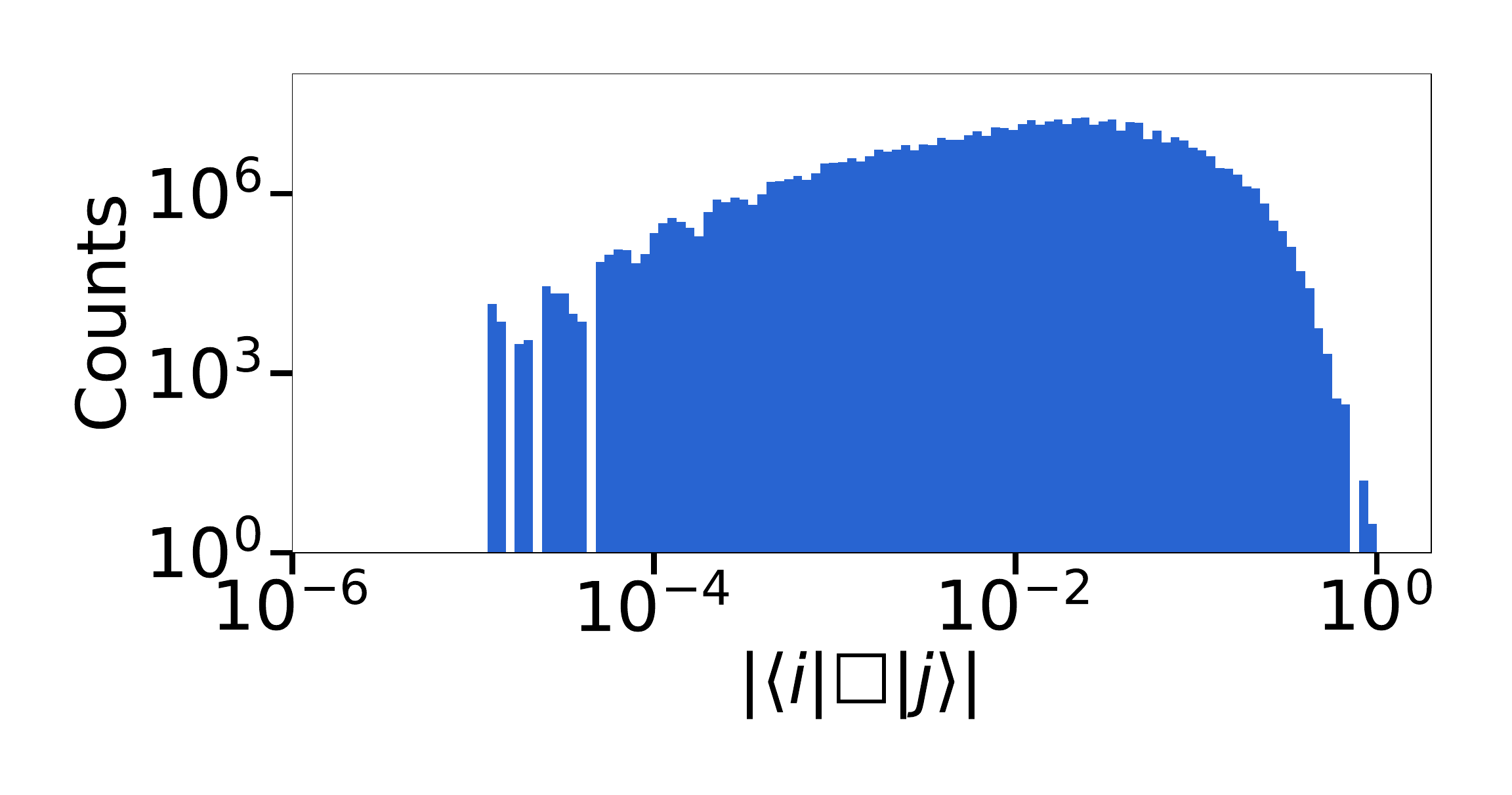}
    \end{subfigure}%
    \caption{Distributions of magnitudes of matrix elements of the plaquette operator  for the $d=2$ (left) and $d=3$ (right) lattices with $T_1$ truncation. }
    \label{figd23distr}
\end{figure}

The plaquette operator matrix elements between physical plaquette states populate distributions. In Figs.~\ref{figd32distr} and~\ref{figd23distr} we show a sampling of these distributions. In the $d=3$ case, the precise matrix elements vary between the three different planes in which a plaquette can lie.\footnote{This is because the $F$-ordering of links needed to define the basis renders the basis non-invariant under discrete rotations. Physical quantities like the expectation value of the electric energy on links will, however, exhibit discrete rotation invariance, and this serves as a nontrivial consistency check on the implementation.} However, the distributions are essentially the same and we show only one representative.

\section{Quantum circuits}
\label{sec:quant sim}

    The time evolution operator for the Kogut-Susskind Hamiltonian is $U(t) = e^{-iHt}$. However, it is difficult and inefficient to create an exact quantum circuit that implements $U(t)$ on a digital quantum computer. Product formulas allow an arbitrary $n$-qubit unitary to be approximated by a product of unitaries. In this work we use the first-order Lie-Trotter approximation. Dividing the time $t$ into $N$ steps of size $\Delta t=t/N$, we have $e^{-iHt} = e^{-iH \Delta t} \cdots e^{-iH \Delta t} + \mathcal{O}(\Delta t^2)$. Each $e^{-iH \Delta t}$ is as difficult to simulate exactly as $e^{-iHt}$, but the short time window $\Delta t$ allows further Trotterization. We decompose  $e^{-iH\Delta t} = \prod_h e^{-ih\Delta t} + \mathcal{O}(\Delta t^2)$, where each of the unitaries $e^{-ih\Delta t}$ is easier to simulate. 

    For the decomposition of each timestep, we use
    \begin{align}
		e^{-iH \Delta t} \approx \prod_{k} e^{-i H_{B,k} \Delta t} \prod_{j} e^{-i H_{E,j} \Delta t}.
		\label{trotter_step}
    \end{align}
    Here each $k$ corresponds to a single Givens rotation on a single plaquette induced by the local magnetic Hamiltonian $\Box_P+\Box^\dagger_P$. Each $j$ corresponds to a term in the Pauli string decomposition of the local electric Hamiltonian on a single link. These decompositions and their circuitizations will be described below.
    
	\subsection{Registers}
	
	First, we encode the degrees of freedom in quantum registers. There is a register dedicated to each of the $N_s$ sites and $N_\ell$ links of the lattice. Link registers encode the irrep states of each link and site registers encode the singlet states of each site in the reduced electric basis. 
    
    We adopt a dense binary encoding of the singlet and irrep states. For example, with 3- and 2-qubit registers for sites and links, respectively, ``000''  encodes some specific singlet state and ``10'' encodes some specific irrep state. In general the site registers contain $n_s = \lceil \log_2(N_s) \rceil$ qubits and the link registers contain $n_r = \lceil \log_2(N_r) \rceil$ qubits, where $N_s$ ($N_r$) is the number of singlets (irreps) of the truncation. Thus, the total number of qubits required is $N_s n_s + N_\ell n_r$.

     We use a particular organization of the quantum registers based on the locations of the plaquettes on a lattice. Each plaquette is identified with a site;  multiple plaquettes are associated with each site for $d>2$. For a given site $s$, choose two unit vectors $e_i$ and $e_j$ with $i<j$. Then the plaquette is defined by $s_1=s$, $\ell_1 = \ell(s,e_i)$, and $\ell_4 = \ell(s,e_j)$. This identification avoids double-counting plaquettes as we sweep over the lattice. Unique site and link registers are then created sequentially, per plaquette, in the order $s_1s_2s_3s_4\ell_1\ell_2\ell_3\ell_4$ as we iterate over all sites. Because the same sites and links may appear in adjacent plaquettes, only the first occurrence of a site/link in this process requires a quantum register. This means that plaquettes which are nearby in the lattice will typically share several link and vertex registers once encoded.
    
	The precise encoding of the singlet and irrep states in register bitstrings is somewhat arbitrary. We choose an irrep encoding based on a lexicographic ordering on the i-weights of the underlying irreps. In the $T_1$ truncation, $(0,0,0) < (1,0,0) < (1,1,0)$, so $(0,0,0) := ``00$'', $(1,0,0) := ``01$'', and $(1,1,0) := ``10$''. Because $N_r=3$ in $T_1$, we require $n_r=2$ qubits per encoded bitstring. The singlet states at each lattice site are completely specified by the irrep values on all the connected half-links along with a zero-indexed positive integer specifying a multiplicity of that singlet. We encode this multiplicity index in an additional register associated with each lattice site. If in a particular geometry and irrep truncation the maximal value which occurs for the zero-indexed multiplicity is $n$, then, each site register will require $\lceil \log_2 (n + 1) \rceil$ qubits. In cases where $n = 0$ such as in the $T_{1}'$ truncation, no site registers are needed. This encoding generalizes straightforwardly to other truncations and dimensions.

	The encoding is used to translate Hamiltonian matrix elements from transitions between singlets and irreps to transitions between bitstrings encoding either single-link or single-plaquette data. The operators $H_{E,j}$ of the electric Hamiltonian are diagonal in the electric basis and remain diagonal in the computational basis after the encoding. Each one-plaquette term $H_{B,k}$ in the magnetic Hamiltonian  acts (non-diagonally) on the states of four site registers, four ``active'' link registers, and $4n$ total ``control'' link registers, $\ket{s_1} \ket{s_2} \ket{s_3} \ket{s_4} \ket{\ell_1} \ket{\ell_2} \ket{\ell_3} \ket{\ell_4}\ket{c_{1,1}}\ldots\ket{c_{4,n}}$.\footnote{The notation $\ket{c_{i,j}}$ denotes the $j$th control link connected to site $s_i$, with some arbitrary convention set to fix the ordering of controls at a particular site.} The number of control link registers per site $n$ depends on the the size of the lattice and its boundary conditions, taking a maximal value of $n = 2(d - 1)$ in $d$ dimensions. The bitstring of the $8 + 4n$ combined registers specifies an electric basis state of a plaquette. Thus the terms in the magnetic Hamiltonian may be thought of as transitions between two particular plaquette bitstrings. The full-lattice Trotter step circuit is then the product of matrix exponentials of these terms.

	Finally, the full-lattice Trotter step circuit is repeatedly appended to obtain the time evolution circuit $U(t)$. This circuit is run on a classical simulator or quantum device and all qubits are measured. Inverting the encoding above, the measured bitstrings can be translated back to singlet and irrep states on the lattice. This allows, for example,  for the calculation of  transition probabilities between electric basis states, as well as chromoelectric expectation values and equal-time correlation functions. We will also employ simple classical post-processing error mitigation to deal with shots falling into unphysical bins. 

	\subsection{Rotation circuits}
	
	The full-lattice Trotter step circuit is a product of $e^{-i H_{E,j} \Delta t}$ circuits and $e^{-i H_{B,k} \Delta t}$ circuits. Because each link register contains $\mathcal{O}(1)$ qubits, we use Pauli string decomposition to construct $e^{-i H_{E,j} \Delta t}$.\footnote{For higher truncations, involving more qubits per link register, it is advantageous to implement a multi-controlled phase gate instead of Pauli decomposition.} However, the set of registers for a plaquette may contain tens of qubits, and $e^{-i H_{B,k} \Delta t}$ is better constructed with multi-controlled unitary  (MCU) gates. In this section we describe our implementation of these circuits. We use the notation $\pi^{0,1}$ for projectors on the $\ket{0}$ and $\ket{1}$ states of a qubit, and $\sigma^\pm$ for the  raising and lowering operators on a qubit.

    \subsubsection{Electric evolution}
    
	Since $H_{E,j}$ is diagonal in the electric basis, it  can be decomposed into a sum of Pauli strings containing only $I$ and $Z$. One way to think of the electric Hamiltonian at a link is in terms of projectors onto each irrep in the truncation. 
	
    For example, consider the term $H_{E,j} = \frac{ g^2 E^2_j}{2a} \pi^0 \pi^1$. 
    Here 
	\begin{equation}
		\pi^0 = \frac{I+Z}{2} \qquad \pi^1 = \frac{I-Z}{2}.
	\end{equation}
    This operator projects onto the irrep encoded by ``10'' with electric Casimir $E^2_j$. Its Pauli decomposition is
	\begin{equation}
		H_{E,j} = \frac{ g^2 E^2_j}{8a}(II - IZ + ZI - ZZ)
	\end{equation}
	The identity string $II$ is discarded because it amounts to a global phase. All $I$-$Z$ strings commute,  so we may write
	\begin{equation}
		e^{-i H_{E,j} \Delta t} = e^{-i \frac{\theta_{IZ}}{2} IZ} e^{-i \frac{\theta_{ZZ}}{2} ZZ} e^{-i \frac{\theta_{ZI}}{2} ZI}
	\end{equation}
	The rotation angles $\theta$ are all $\pm \frac{ g^2E^2_j}{4a} \Delta t$. Strings that contain one $Z$ are simply an $R_Z(\theta)$ gate on that qubit. Strings that contain more than one $Z$ require a parity-counting scheme, e.g.:
    \begin{equation}
        e^{-i \frac{\theta}{2} ZZZ} =
		\begin{quantikz}[align equals at=2, row sep={24pt,between origins}]
			\lstick{$q_2$} & \ctrl{2} & & & & \ctrl{2} & \\
			\lstick{$q_1$} & & \ctrl{1} & & \ctrl{1} & & \\
			\lstick{$q_0$} & \targ{} & \targ{} & \gate{R_Z(\theta)} & \targ{} & \targ{} &
		\end{quantikz}
    \end{equation}

    \subsubsection{Magnetic evolution}
    \label{sec:mag evol}

    To Trotterize the magnetic timestep at a plaquette, we expand the magnetic plaquette Hamiltonian in a sum of two-state transitions. In this decomposition, each operator $H_{B,k}$ is proportional to a (symmetrized) string of projector and ladder operators, e.g.
    \begin{equation}
        \pi^1 \sigma^+ \sigma^- \pi^0 + \text{h.c.} \equiv \ketbra{1100}{1010} + \text{h.c.}
    \end{equation}
    In this form it is clear that each term in the decomposition encodes a transition between two plaquette states $P_i$ and $P_f$. Thus each factor $e^{-i H_{B,k} \Delta t}$ in the Trotterization is a Givens rotation\footnote{ A Givens rotation rotates two components of a vector into each other. $e^{-i H_{B,k} \Delta t}$ rotates two computational basis states, corresponding to a complex Givens rotation on the computational Hilbert space. This is simply the most obvious way to implement a unitary given by an
exponential of a matrix that has no particular structure. This approach is often used to implement magnetic evolution in an electric basis, see e.g.~\cite{Ciavarella:2021nmj,Ciavarella:2022zhe}. An early adopter of Givens rotations
was Jacobi, who used them in an iterative algorithm to diagonalize matrices~\cite{Jacobi+1846+51+94}.} with a rotation angle proportional to $\bra{P_f} \Box + \Box^\dag \ket{P_i}$.

To construct the circuit for $e^{-i H_{B,k} \Delta t}$, we diagonalize $H_{B,k}$ using $CX$ gates. 
To do this, we observe that terms in the magnetic Hamiltonian can be grouped into families of mutually commuting operators. A convenient choice is given by the $LP$ \emph{families} ($L$ for ``ladder'' and $P$ for ``projector''). 
An $LP$ family is designated by a choice of $L$ or $P$ for each qubit; e.g., $LLPLPPL \dots$, etc. 
On $n$ qubits there are $2^n$ commuting operators per $LP$ family. Each magnetic Hamiltonian term can be grouped into the $LP$ family corresponding to its arrangement of ladder and projector operators. For instance, $\pi^1 \sigma^- \pi^1 + \text{h.c.}$  belongs to the $PLP$ family. Then, for each $LP$ family, there is a  circuit $V$ that transforms all $H_{B,k}$ in the family  into operators comprised of $n-1$ projectors and a rotation on a single \emph{pivot qubit} $q'$.\footnote{$V$ and $q'$ depend on the $LP$ family, but this dependence is suppressed in the notation.}
    
\begin{figure}[ht]
    \centering
    $e^{-i\frac{\theta}{2}(\pi^1 \sigma^+ \sigma^- \pi^0 + \text{H.c.})} = $
    \begin{quantikz}[align equals at=2.5, row sep={24pt,between origins}]
        \lstick{$q_3$} & & \ctrl{2} & & \rstick{$(P)$} \\
        \lstick{$q_2$} & \ctrl{1} & \gate{R_X(\theta)} & \ctrl{1} & \rstick{$(L)$} \\
        \lstick{$q_1$} & \targ{} & \control{} & \targ{} & \rstick{$(L)$} \\
        \lstick{$q_0$} & & \octrl{-1} & & \rstick{$(P)$}
    \end{quantikz}
    \caption{Givens rotation circuit for magnetic evolution. The $CX$ gates transform the $q_2$ ladder operator into a projector. Then $e^{-i\frac{\theta}{2} \pi^1 X \pi^1 \pi^0}$ is a multi-controlled $R_X(\theta)$ gate. Note that the magnetic Hamiltonian term shown here belongs to the $PLLP$ family and $q_2$ is $q'$.}
    \label{givens_ex}
\end{figure}
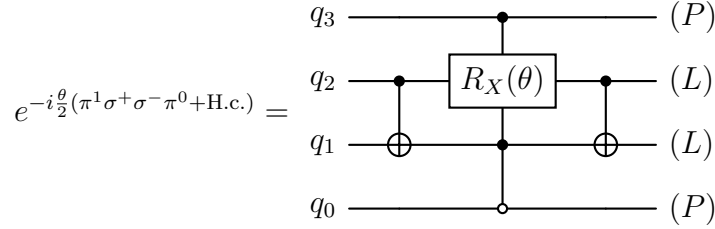

For each $H_{B,k}$, we fix a convention where the pivot qubit $q'$ is identified as the first qubit acted on by $\sigma^\pm$ (i.e.~the first $L$ in the $LP$ family), when reading $H_{B,k}$ from left to right.  The circuit $V$ can then be read off of the $LP$ family in the following way. For each qubit after $q'$, append a $CX$ gate controlled on $q'$ and targeted to each other qubit on which there is an $L$. These entangling $CX$ gates cause the rotation of the pivot to propagate to all other  $L$-qubits. Then $H_{B,k} \to V \tilde{H}_{B,k} V^\dag$, where $\tilde{H}_{B,k}$ is a string of projectors and an $X$ gate on $q'$ (e.g.  $\pi^1 X \pi^1 \pi^0$). Thus $e^{-i H_{B,k} \Delta t} = V e^{-i \tilde{H}_{B,k} \Delta t} V^\dag$, where  $e^{-i \tilde{H}_{B,k} \Delta t}$ is a multi-controlled $R_X(\theta)$ gate. An example is shown in Fig.~\ref{givens_ex}.
The rotation angle, $\theta = \frac{2}{2ag^2} \bra{P_f} \Box + \Box^\dag \ket{P_i} \Delta t$, is classically precomputed.  Because multi-controlled $R_X(\theta)$ gates are not native on most current quantum hardware, it is useful to decompose them into multi-controlled $X$ gates, which can then be decomposed into $CX$s:
\begin{equation}
    \begin{quantikz}[align equals at=2.5, row sep={24pt,between origins}]
			& \ctrl{2} & \\
			& \gate{R_X(\theta)} & \\
			& \control{} & \\
			& \octrl{-2} &
		\end{quantikz} = 
		\begin{quantikz}[align equals at=2.5, row sep={24pt,between origins}]
			& & & \ctrl{1} & & \ctrl{1} & & \\
			& \gate{R_Z(-\frac{\pi}{2})} & \gate{R_Y(-\frac{\theta}{2})} & \gate{X} & \gate{R_Y(\frac{\theta}{2})} & \gate{X} & \gate{R_Z(\frac{\pi}{2})} & \\
                & & & \control{} & & \control{} & & \\
			& & & \octrl{-2} & & \octrl{-2} & &
		\end{quantikz}
\end{equation}

    \subsection{Resources and optimization}
    \label{sec:resource and optimizations}

        The bulk of the gate depth in the Trotterized circuit comes from multi-controlled $CX$ gates ($C^nX$) which generate the multi-controlled rotations ($MCR$). While native implementations of multi-controlled unitaries are a focus of research interest \cite{covey,limeng}, it is currently more feasible to decompose them into single- and double-qubit unitaries. Of particular interest are the double-unitary $CX$ or $CNOT$ gate and the non-Clifford unitary $T$, and as the number of matrix elements, plaquettes, and qubits per plaquette register grow, the $CX$ and $T$ gate costs grow as well. In an effort to reduce the depth of $e^{-iH\Delta t}$, we perform several optimizations.
	
	\subsubsection{Gray code ordering}
	
	In both electric and magnetic evolution circuits, there are $CX$ gates that surround a rotation. It is possible to reduce these $CX$s by ordering the local Hamiltonian operators in a Gray code \cite{Nielsen_Chuang_2010} fashion described below.
	
	In the electric evolution, each $H_{E,j}$ contains a sum of Pauli strings. It is always possible to reorder these Pauli strings such that their sequence returns a Gray code. In such a sequence, neighboring terms differ by one Pauli operator. For example, $(IZ,ZZ,ZI)$ gives a Gray code sequence because $IZ$ ($ZI$) differs from $ZZ$ only on the left (right) operator. Another example of a Gray code sequence is $(IIZ,IZZ,IZI,ZZI,ZZZ,ZIZ,ZII)$. In the matrix exponential, $e^{-i H_{E,j} \Delta t}$, a Gray code ordering will ensure that at most 1 $CX$ is necessary between each $R_Z(\theta)$ gate.

    Generally, the magnetic Hamiltonian will contain terms from many different $LP$ families. The $CX$s present in each $e^{-i H_{B,k} \Delta t}$ appear in relation to the $L$ operators of their respective $LP$ family. This suggests an optimization done by ordering the $LP$ families in Gray code. For example, $(PLL,PLP,LLP,LPP)$ is a Gray code sequence of $LP$ families. Of course, the number of $LP$ families in the magnetic Hamiltonian can be much smaller than $2^n$, but this is an optimal ordering of the families.

    \subsubsection{Control pruning}
	
	One of the simplest lattices is a single plaquette with open boundary conditions. In the $T_1$ irrep truncation, this lattice requires $n_r=2$ qubits per link and no site qubits. Therefore, encoding this lattice requires 8 qubits, and the magnetic evolution circuit na\"ively contains $C^7X$ gates. However, there are only three physical plaquette states. The 7 controls on the $C^7X$ are ensuring that only two out of the $2^8=256$ possible basis states in the computational Hilbert space are rotated into each other, even though 253 of these states will never appear in a noiseless simulation. Typically, many of the 7 controls per $C^7X$  can be removed, or ``pruned," without changing the action of the unitary operator on the physical subspace. Pruning in the classical circuit preparation stage can substantially lower the $CX$ and $T$ gate cost of the magnetic evolution, once the MCUs are decomposed into one and two qubit gates.
	
	The control pruning algorithm requires knowledge of the set $\mathcal{P}$ of  gauge-invariant plaquette states present in the given truncation scheme, the bitstrings $P_i$ and $P_f$ involved in the Givens rotation, and the $LP$ family corresponding to the rotation.\footnote{This data is automatically collected during the classical construction of $\Box$ matrix elements.} The output is the set of qubits $Q$ in the Givens rotation that must be controlled on. The algorithm is  as follows:

    \begin{enumerate}
		\item Choose a representative $P$ from the Givens rotation bitstrings (either $P_i$ or $P_f$.) Identify the pivot qubit $q'$ associated with $P$ (see Sec.~\ref{sec:mag evol} for the definition of the pivot.) Because the $R_X(\theta)$ gate of the Givens rotation is applied to $q'$, $q'$ cannot be in $Q$, but any other qubit associated with the plaquette may appear in $Q$.
		\item Using the diagonalizing circuit $V$ described in Sec.~\ref{sec:mag evol}, translate $\mathcal{P}$ into $\tilde{\mathcal{P}}$, the set of physical plaquette states after applying $V$. For example, if the family is $LLPP$, then $V$ takes $1100\rightarrow 1000$, $0000\rightarrow 0000$, etc. Similarly, relabel the representative $P \to \tilde{P}$. (This step is necessary because the multi-controlled $R_X(\theta)$ in a Givens rotation circuit operates on the states in $\tilde{\mathcal{P}}$ rather than $\mathcal{P}$.)
		\item Iterate once through $\tilde{\mathcal{P}}$,  identifying any bitstrings (excluding the pivot qubit $q'$) with Hamming distance one from $\tilde{P}$. These qubits are a subset of the necessary control qubits; without them, these bitstrings would be erroneously rotated by the $MCRX$. Collect these qubits into the set $Q$, and eliminate all bitstrings from $\tilde{\mathcal{P}}$ that differ from $\tilde{P}$ at any qubits in $Q$: these strings are vetoed by controls on the $Q$.\footnote{While this step is not strictly necessary, it reduces the number of loops required in the final steps of the algorithm.}
		\item\label{alg:iter step} Iterate through the remaining bitstrings of $\tilde{\mathcal{P}}$. For each bitstring, populate the list of qubits (excluding the pivot) at which the bitstring differs from $\tilde{P}$. At the end of this iteration, find the qubit $q$ that appeared most frequently in these lists. (If there are multiple, choose  one.) Append $q$ to $Q$, and eliminate from $\tilde{\mathcal{P}}$ all bitstrings that differ from $\tilde{P}$ at $q$.
		\item Repeat step \ref{alg:iter step} until $\tilde{\mathcal{P}}$ only contains $\tilde{P_i}$ and $\tilde{P_f}$.
	\end{enumerate}
	The algorithm terminates with the set $Q$, the set of control qubits necessary to distinguish $P_i$ and $P_f$ from all other states in $\mathcal{P}$. The number of controls scales logarithmically with the number of plaquette states: $|Q| \approx \lceil \log_2(|\mathcal{P}|) \rceil$.

    \subsubsection{Control fusion}

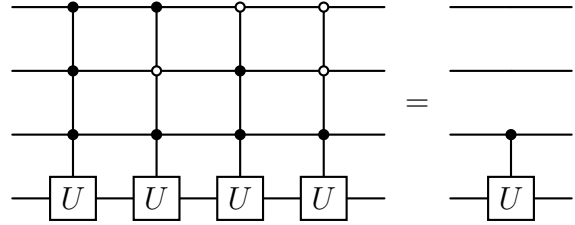
\begin{figure}[ht]
		\centering
		\begin{quantikz}[align equals at=2.5, row sep={24pt,between origins}]
			& \ctrl{3} & \ctrl{3} & \octrl{3} & \octrl{3} & \\
			& \control{} & \ocontrol{} & \control{} & \ocontrol{} & \\
			& \control{} & \control{} & \control{} & \control{} & \\
			& \gate{U} & \gate{U} & \gate{U} & \gate{U} &
		\end{quantikz} $ = $
		\begin{quantikz}[align equals at=2.5, row sep={24pt,between origins}]
			& & \\
			& & \\
			& \ctrl{1} & \\
			& \gate{U} &
		\end{quantikz}
		\caption{Example of a fusion of multi-controlled unitaries.}
		\label{ctrl_combine}
	\end{figure}

    A final optimization of the multi-controlled unitaries is possible for a given $LP$ family. If there are multiple terms $H_{B,k}$ of the magnetic Hamiltonian that correspond to the same $LP$ family and they possess the same $\bra{P_f} \Box + \Box^\dag \ket{P_i}$ coefficient, then it may be possible to fuse these terms in powers of two. This is due to the property shown in fig.~\ref{ctrl_combine}.
	
\subsubsection{Quantum resource estimates}

Here we provide rough estimates for the quantum resources required to run simulations, in terms of $CX$ and $T$ gate counts. 
\begin{itemize}
    \item For a $C^nX$ gate with $n$ controls, we quote the result of \cite{zindorf2024efficientimplementationmulticontrolledquantum}, which found a decomposition in terms of $12n-32$ $CX$ gates and $16n-48$ $T$ gates for $n\geq 6$, without ancillas. With ancillas, we quote the result of \cite{maslov}, which found a decomposition in terms of $6n-18$ $CX$ gates and $8n-25$ $T$ gates.
    \item If there are $N_M$ terms in the magnetic Hamiltonian, then there are $2N_M$ $C^nX$ gates in a Trotter circuit without optimization.\footnote{Note that depending on implementation, one may be able to decompose the multi-controlled rotation directly into $CX$ rather than into two $MCX$ \cite{maslov}.} The $CX$s between each term add no more than $2nN_M$ $CX$s. Neglecting the electric evolution which contributes a small number of $CX$ and further optimizations we obtain the estimates tabulated in Table~\ref{generalgatetbl}.
    
    \item Let $N_\Box$ be the number of physical plaquette states in the simulation. Control pruning causes each $C^nX$ to be controlled on order $\lceil \log_2(N_\Box) \rceil$ qubits. If control pruning is the only optimization used, then  $n$ in the points above can be replaced by $\lceil \log_2(N_\Box) \rceil$.
    \item If control fusion and Gray code orderings are used, then the resources can vary  further. Control fusion reduces both $N_M$ and $n$. Gray code orderings linearly reduce the $CX$ cost.

\end{itemize}
    \begin{table}[h!]
        \centering
        \begin{tabular}{|c|c|c|}
            \hline
            Gate & $CX$ & $T$ \\
            \hline
            w/o Ancillas & $2N_M(13n-32)$ & $2N_M(16n-48)$ \\
            \hline
            w/ Ancillas & $2N_M(7n-18)$ & $2 N_M(8n-25)$ \\
            \hline
        \end{tabular}
        \caption{General gate cost estimates per plaquette per Trotter step with no optimizations. $n+1$ is the number of qubits needed to encode a plaquette state.}
        \label{generalgatetbl}
    \end{table}

\begin{table}[ht]
\centering
 \resizebox{0.85\textwidth}{!}{
\begin{tabular}{l|ll|rrrrcc|}
\cline{2-9}
 &Lattice& $T_i$ & Trans & Qubits & $MCX$ & Ctrl & $CX$ & $T$ \\ \hline
\rowcolor[HTML]{EFEFEF} 
\multicolumn{1}{|l|}{\cellcolor[HTML]{EFEFEF}Unopt} & 2-plaq & $T_1$ & 27 & 12 & 108 & 11 & $1.20\times10^4$ & $1.38\times 10^4$ \\
\multicolumn{1}{|l|}{Opt} & & &  & 15 & 108 & 4 & $2.80\times10^3$ & $3.10\times10^3$ \\ \hline
 & \cellcolor[HTML]{EFEFEF}3-plaq & \cellcolor[HTML]{EFEFEF}$T_1$ & \cellcolor[HTML]{EFEFEF}81 & \cellcolor[HTML]{EFEFEF}18 & \cellcolor[HTML]{EFEFEF}486 & \cellcolor[HTML]{EFEFEF}15 & \cellcolor[HTML]{EFEFEF}$7.92\times10^4$ & \cellcolor[HTML]{EFEFEF}$9.33\times10^4$ \\
 & & &  & 23 & 486 & 4 & $1.25\times10^4$ & $1.40\times10^4$ \\ \cline{2-9} 
 & \cellcolor[HTML]{EFEFEF}5-plaq & \cellcolor[HTML]{EFEFEF}$T_1$ & \cellcolor[HTML]{EFEFEF}81 & \cellcolor[HTML]{EFEFEF}30 & \cellcolor[HTML]{EFEFEF}810 & \cellcolor[HTML]{EFEFEF}15 & \cellcolor[HTML]{EFEFEF}$1.32\times10^5$ & \cellcolor[HTML]{EFEFEF}$1.56\times10^5$ \\
 &  & &  & 35 & 810 & 4 & $2.05\times10^4$ & $2.30\times10^4$ \\ \cline{2-9} 
 & \cellcolor[HTML]{EFEFEF}2-plaq & \cellcolor[HTML]{EFEFEF}$T_2$ & \cellcolor[HTML]{EFEFEF}3798 & \cellcolor[HTML]{EFEFEF}22 & \cellcolor[HTML]{EFEFEF}15192 & \cellcolor[HTML]{EFEFEF}21 & \cellcolor[HTML]{EFEFEF}$3.66\times10^6$ & \cellcolor[HTML]{EFEFEF}$4.38\times10^6$ \\
 & & &  & 33 & 14912 & 9 & $8.10\times10^5$ & $9.40\times10^5$ \\ \cline{2-9} 
 & \cellcolor[HTML]{EFEFEF}$2\times2$ & \cellcolor[HTML]{EFEFEF}$T_1$ & \cellcolor[HTML]{EFEFEF}835 & \cellcolor[HTML]{EFEFEF}20 & \cellcolor[HTML]{EFEFEF}6680 & \cellcolor[HTML]{EFEFEF}19 & \cellcolor[HTML]{EFEFEF}$1.44\times10^6$ & \cellcolor[HTML]{EFEFEF}$1.71\times10^6$ \\
 & & &  & 27 & 5104 & 7 & $2.20\times10^5$ & $2.50\times10^5$ \\ \cline{2-9} 
 & \cellcolor[HTML]{EFEFEF}$3\times3$ & \cellcolor[HTML]{EFEFEF}$T_1'$ & \cellcolor[HTML]{EFEFEF}83 & \cellcolor[HTML]{EFEFEF}36 & \cellcolor[HTML]{EFEFEF}1494 & \cellcolor[HTML]{EFEFEF}23 & \cellcolor[HTML]{EFEFEF}$3.99\times10^5$ & \cellcolor[HTML]{EFEFEF}$4.78\times10^5$ \\
 & & &  & 46 & 1494 & 8 & $7.25\times10^4$ & $8.40\times10^4$ \\ \cline{2-9} 
 & \cellcolor[HTML]{EFEFEF}1-cube & \cellcolor[HTML]{EFEFEF}$T_1$ & \cellcolor[HTML]{EFEFEF}81 & \cellcolor[HTML]{EFEFEF}24 & \cellcolor[HTML]{EFEFEF}972 & \cellcolor[HTML]{EFEFEF}15 & \cellcolor[HTML]{EFEFEF}$1.58\times 10^5$ & \cellcolor[HTML]{EFEFEF}$1.87\times 10^5$ \\
 & & &  & 29 & 972 & 4 & $2.45\times10^4$ & $2.75\times10^4$ \\ \cline{2-9} 
 & \cellcolor[HTML]{EFEFEF}$2\times2\times2$ &  \cellcolor[HTML]{EFEFEF}$T_1'$ & \cellcolor[HTML]{EFEFEF}127 & \cellcolor[HTML]{EFEFEF}48 & \cellcolor[HTML]{EFEFEF}6096 & \cellcolor[HTML]{EFEFEF}31 & \cellcolor[HTML]{EFEFEF}$2.26\times 10^6$ & \cellcolor[HTML]{EFEFEF}$2.73\times 10^6$ \\
 & & &  & 66 & 6096 & 8 & $2.90\times10^5$ & $3.40\times10^5$ \\ \cline{2-9} 
 & \cellcolor[HTML]{EFEFEF}$3\times3\times3$ &  \cellcolor[HTML]{EFEFEF}$T_1'$ & \cellcolor[HTML]{EFEFEF}707 & \cellcolor[HTML]{EFEFEF}162 & \cellcolor[HTML]{EFEFEF}114534 & \cellcolor[HTML]{EFEFEF}39 & \cellcolor[HTML]{EFEFEF}$5.44\times10^7$ & \cellcolor[HTML]{EFEFEF}$6.60\times10^7$ \\
 & & &  & 180 & 114534 & 9 & $6.19\times10^6$ & $7.22\times10^6$ \\ \cline{2-9} 
\end{tabular}
}
\caption{
    Gate cost comparison between unoptimized and optimized circuits. ``Trans(itions)" refers to the number of $MCR$ per trotter step per plaquette (equivalently, it is the number of nonzero matrix elements of $\Box$). ``Ctrl" is the average number of controls per $MCX$. ``Unopt" means that each $MCR$ is directly transpiled into $CX$ and $T$ and then combined. ``Opt" means that each $MCR$ has been pruned, then Gray-ordered, then fused, then implemented using ancilla qubits and the v-chain method \cite{oldshor}. $n$-plaq are $d=3/2$ lattices of length $n$ with PBC (Fig.~\ref{twoplaq}). $n\times n$ are $d=2$ lattices of size $n^2$ with PBC (Fig.~\ref{d2lat}). 1-cube is a $d=3$ lattice with open boundary conditions (Fig.~\ref{cubeobc}). $n\times n\times n$ are lattices of size $n^3$ with PBC. All-to-all connectivity is assumed in $CX$ counts. Table made with \cite{latex_table}.}
    \label{gatestable}
\end{table}

\begin{figure}[h!]
    \centering
        \includegraphics[width=0.7\linewidth]{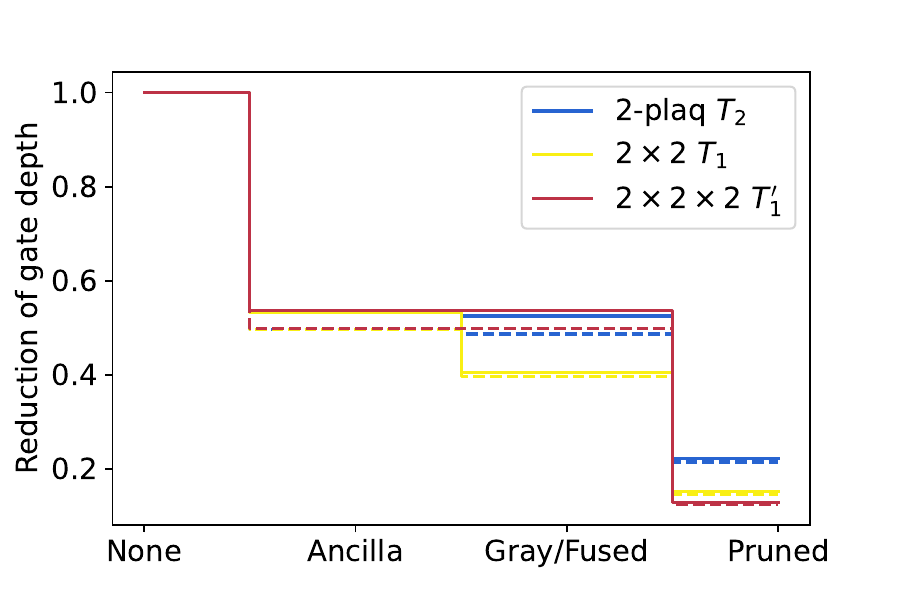}
    \caption{Reduction of gate depth for $CX$ count (solid) and $T$ count (dashed) for three of the lattices described in Table~\ref{gatestable}.}
    \label{gate_reduction}
\end{figure}

Table~\ref{gatestable} summarizes gate counts and resource requirements  for a variety of small lattices, based on explicit circuit construction, and illustrates the improvements afforded by the pruning and fusion optimizations. Simulation results for several of these lattices are reported in the next section. As seen also in Fig.~\ref{gate_reduction}, using ancilla in the $MCX$ decomposition reduces gate depth by about half for any lattice. Control pruning has a larger impact for lattices with a larger control scheme, with almost a five-fold reduction for the $2\times2\times2$ lattice. The effect of pruning can be seen in the decrease of the number of control qubits in Table ~\ref{gatestable}. Observe that Gray-ordering and control fusion impact lattices with a large number of transitions relative to the size of their control scheme, but have negligible impact for lattices with strong truncations on the number of physical states. The effects of control fusion can be seen in the decrease of $MCX$ after optimization.

\section{Simulations}
\label{results}
In this section we report results for noiseless simulations and noisy emulations of Quantinuum systems  in $d=3/2$, $d=2$, and $d=3$. Throughout, we begin in the electric vacuum state and measure the expectation value of the electric field $\langle E^2\rangle$ averaged over the lattice.  

Due to the rapid growth in the number of physical plaquette states and magnetic matrix elements with dimension and truncation, it is essential to understand which matrix elements contribute significantly to a given simulation. Matrix elements which contribute insignificantly to a noiseless simulation still contribute noise on a real device, and cost computation time. Therefore, a main focus in this section will be to  develop assessments of significance (``scores") for the pre-computed matrix elements, and test additional truncations based on these scores.

\subsection{$d=3/2$}

\begin{figure}[ht]
    \centering
        \includegraphics[width=0.2\linewidth]{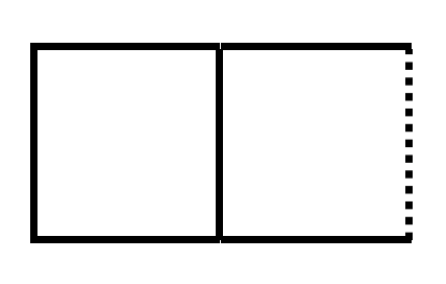}
    \caption{The 2-plaquette lattice with PBCs.}
    \label{twoplaq}
\end{figure}

\begin{figure}[ht]
    \centering
    \includegraphics[width=\linewidth]{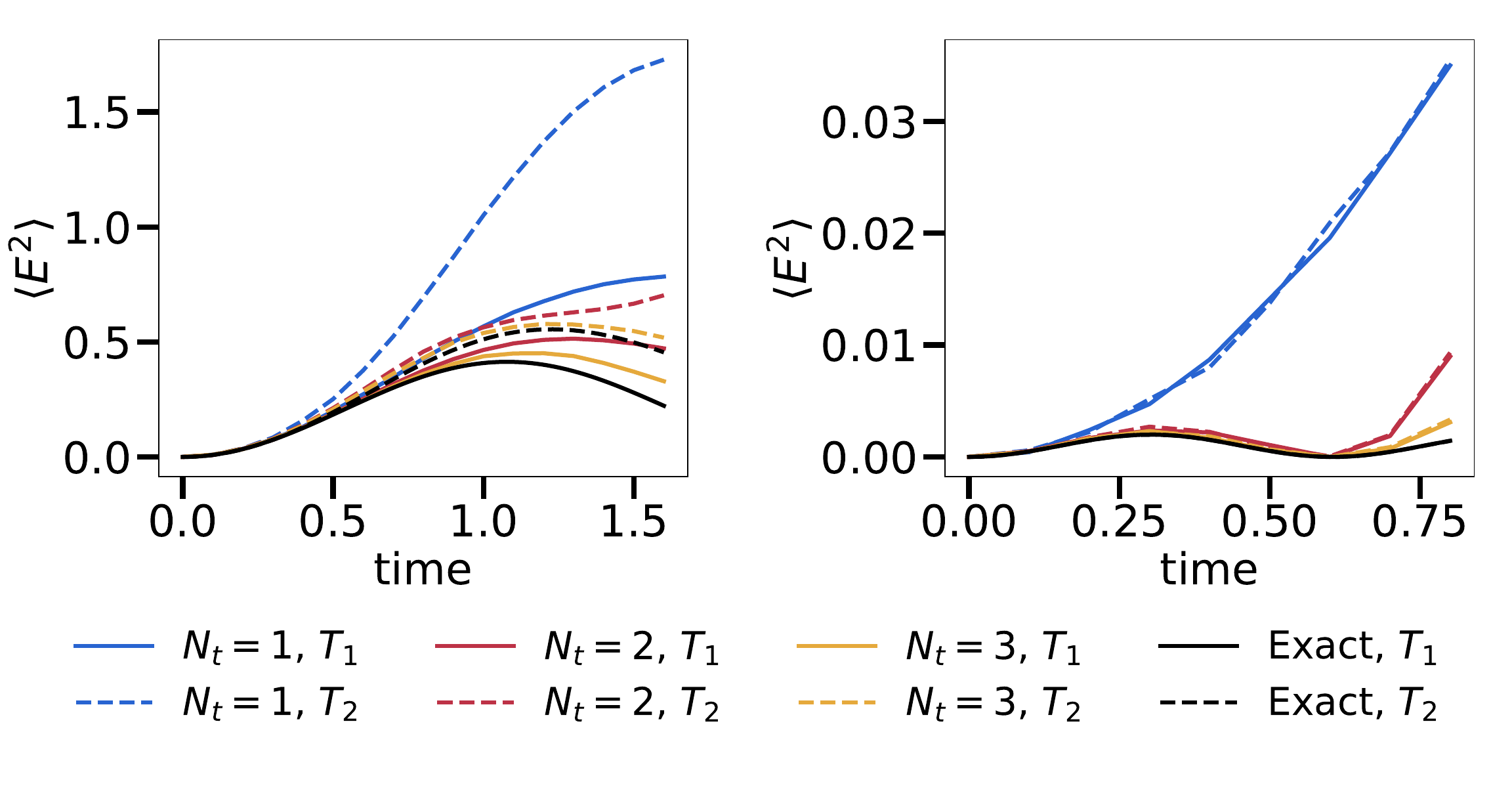}
    \caption{Time evolution of the link-averaged electric Casimir operator, starting from the electric vacuum on the $d=3/2$ two-plaquette lattice. Different curves correspond to different numbers of Trotter steps $N_t$ and irrep truncations $T_{1}$ and $T_2$. Left: $g=1$. Right: $g=2$. ``Exact" curves refer to results obtained by exact diagonalization. For larger coupling, the impact of the irrep truncation is suppressed.}
    \label{figd32T1T2}
\end{figure}

The ``line of plaquettes" lattice, in the local basis, with $T_1$ truncation,  two plaquettes, and periodic boundary conditions along the line, was studied in~\cite{Ciavarella:2021nmj}. The lattice is shown in  Fig.~\ref{twoplaq}. Fig.~\ref{figd32T1T2} compares the Trotter evolution of the electric vacuum in the $T_1$ and $T_2$ truncations on this lattice at couplings $g=1$ and $g=2$.\footnote{We have checked that the matrix elements used for the $T_1$ truncation in $d=3/2$ agree with those of~\cite{Ciavarella:2021nmj}. The left-hand plot of Fig.~\ref{figd32T1T2} agrees with Figs 12 (14) of~\cite{Ciavarella:2021nmj}, after rescaling by 3 (6) to match the whole-lattice electric energy (whole-lattice $\langle E^2\rangle$).}  We observe that including two-index tensor irreps has a much smaller impact at stronger coupling, reflecting larger gaps.\footnote{
Recent work~\cite{ciavarella2025generichilbertspacefragmentation} has studied a general phenomenon of ``fragmentation" of the lattice gauge theory Hilbert space, where transitions between sectors of significantly different electric flux are effectively decoupled at larger $g$. This is a distinct phenomenon with similar effects.}

On this lattice, with the $T_1$ irrep truncation, the plaquette operator can induce 27 different transitions between physical plaquette states. To assess the relative importance of each transition, we assign a score:
\begin{align}
    {\rm Score}(i\leftrightarrow j, t) = \text{max}\left(|\langle j|(\Box + \Box^{\dagger})|i\rangle| P(i,t) \, dt, \, \, |\langle j|(\Box + \Box^{\dagger})|i\rangle| P(j,t) \, dt \right)
    \label{eq:score}
\end{align}
Here $P(i,t)$ is the probability that a given plaquette is in physical state $i$ at time $t$. Thus the score measures the importance of an $i \leftrightarrow j$ transition at a given time, weighing both the amplitude of the transition and the probability of being in one of the states in the transition. To get an accurate sample of the significant transitions involved in time evolution up to some time $T$, we sample scores at several values of $0 < t < T$ and take the maximum score for each matrix element over the sample.

\begin{figure}[ht]
    \centering
    \begin{subfigure}[t]{0.5\linewidth}
        \centering
        \includegraphics[width=\linewidth]{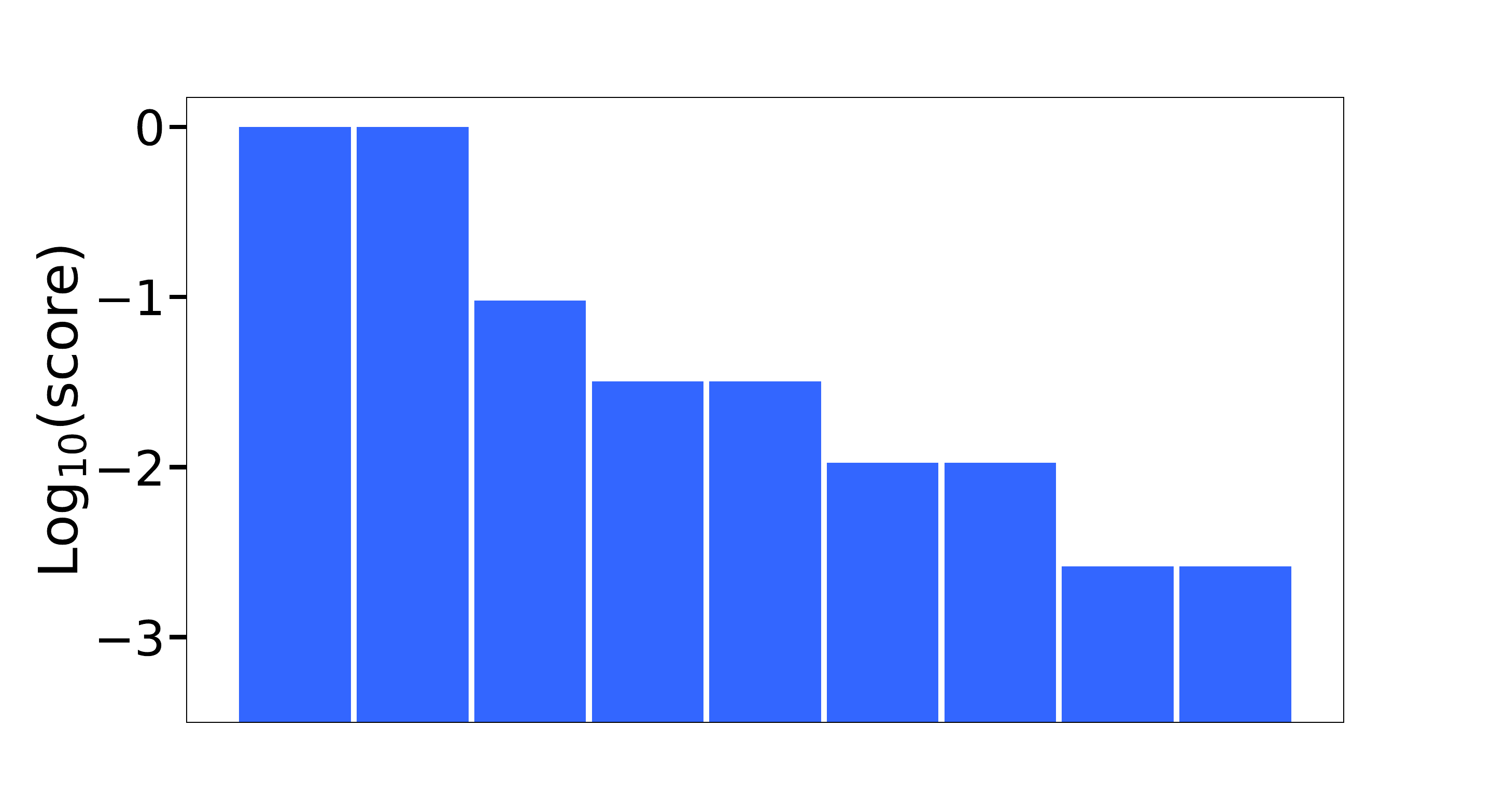}
    \end{subfigure}%
    ~
    \begin{subfigure}[t]{0.5\linewidth}
        \centering
        \includegraphics[width=\linewidth]{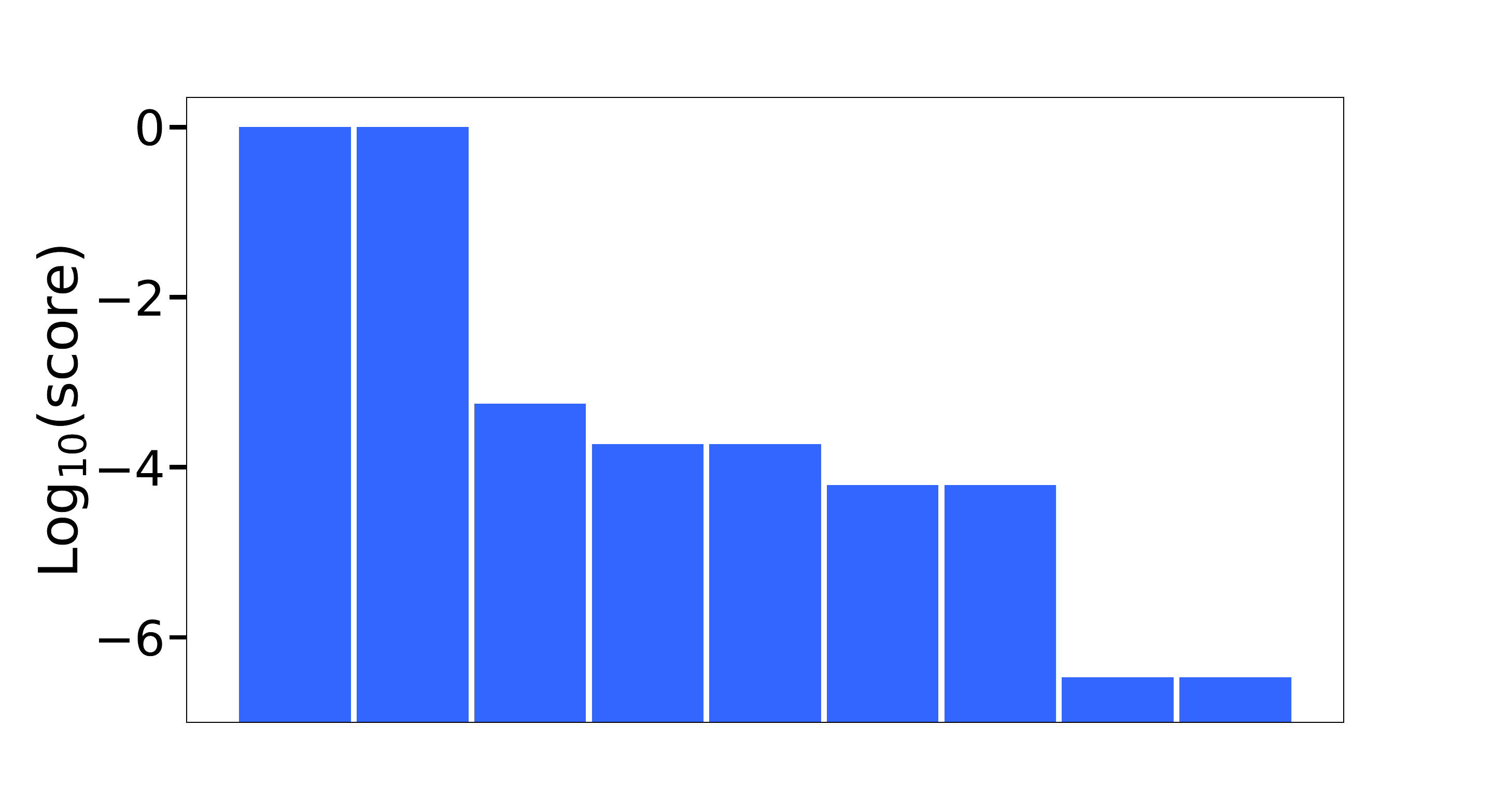}
    \end{subfigure}%
    \caption{Matrix element scores for the $d=3/2$ two-plaquette lattice with   $T_1$ irrep truncation, and $g=1$ (left) and $g=2$ (right).}
    \label{figd32biggestscores}
\end{figure}

\begin{figure}[ht]
    \centering
    \includegraphics[width=\linewidth]{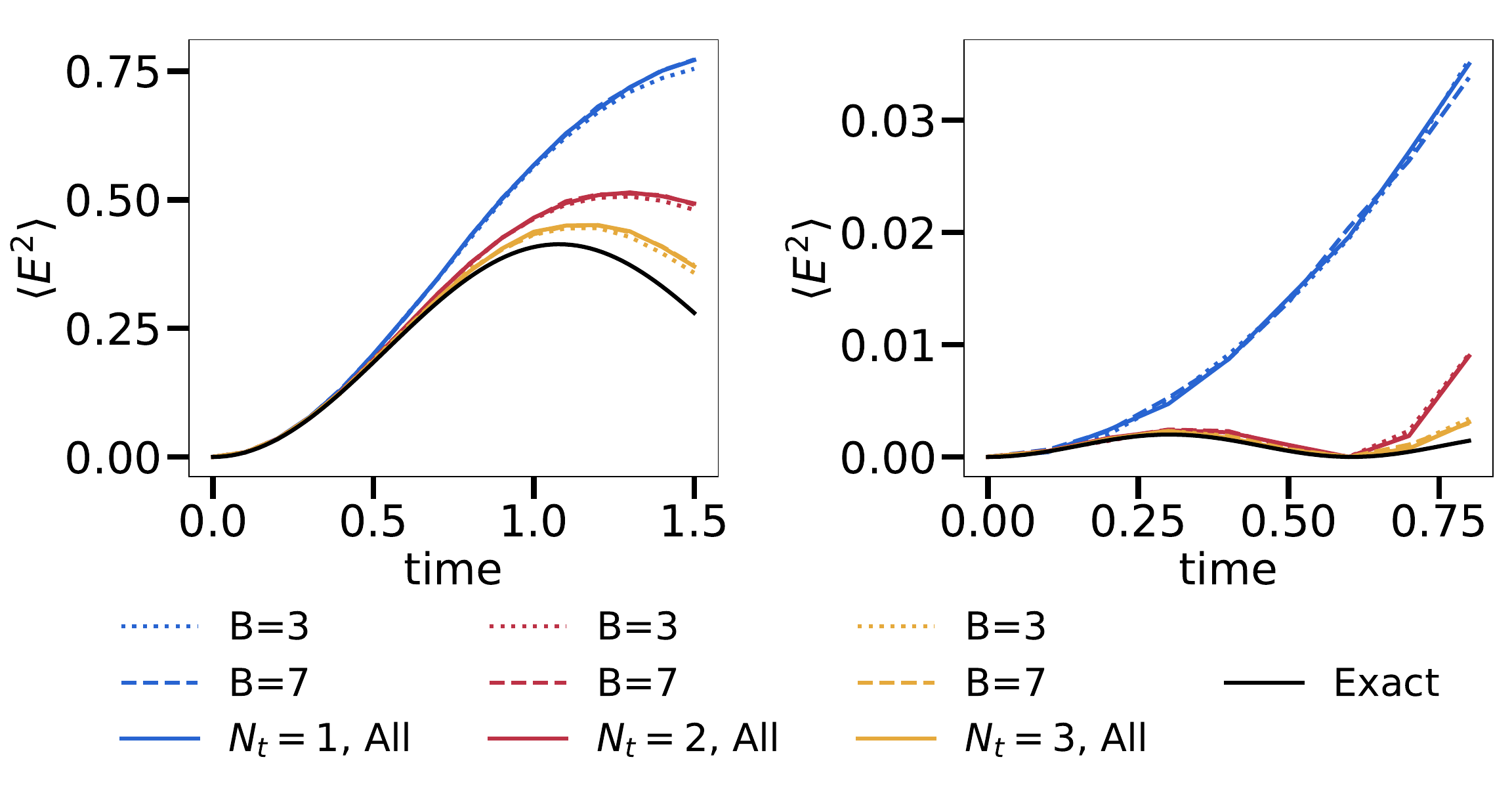}
    \caption{Time evolution of the link-averaged electric Casimir operator, starting from the electric vacuum on the $d=3/2$ two-plaquette lattice in the $T_1$ irrep truncation. Different curves correspond to different numbers of Trotter steps $N_t$ and truncations on the magnetic matrix element scores shown in Fig.~\ref{figd32biggestscores}; only the  elements with the $B$ largest scores are kept in the truncation.  Left: $g=1$. Right: $g=2$.}
    \label{figd32scoringanalysis}
\end{figure}

Fig.~\ref{figd32biggestscores} shows the  distribution of (nonzero) matrix element scores, for $g \! = \! 1$ and $g \! = \! 2$ and $T \! = \! 3$, sampling scores over five equally spaced times. Each bar represents a matrix element. Notably, only 9 out of 27 matrix elements actually participate; others describe transitions between physical states that cannot be reached by time evolution from the electric vacuum. Among the 9, most are numerically insignificant, and the gap between scores grows with $g$. Fig.~\ref{figd32scoringanalysis} shows the effect of truncating transitions based on score. In this case, keeping only the first 3 matrix elements---transitions away from the vacuum and a transition involving the first set of excited states---still gives an excellent approximation to the evolution with all matrix elements. 

We emphasize that  matrix element truncation based on the score introduced above is not a scalable algorithm, since it takes input from classical simulations that are limited to $\lesssim 40$ qubits. It is also dependent on the initial state and duration of evolution, and may have different performance for different observables. (For observables $\mathcal{O}$ that are diagonal in the electric basis, a variant of the algorithm is to weight the scores  by $\mathcal{O}(i)$ or $\mathcal{O}(j)$ for an $i \leftrightarrow j$ transition.) We introduce the score cut only to illustrate the point that not all matrix elements contribute equally importantly to the evolution of a given state. Exploiting this point is likely to be essential in scaling simulations to larger irrep truncations and dimensions, where the number of physical plaquette states and transitions grows rapidly. A more elegant and scalable version of the algorithm would attempt to construct a score without requiring full classical simulations. We will not construct such an algorithm here.\footnote{Previous works have examined cuts on matrix element magnitudes alone, see e.g.~\cite{Kane:2022ejm}. Ideally, a more scalable algorithm will also account for the ``cost" of getting to the states $i$ or $j$, as the score does.} Below, we will assess the same scoring and truncation algorithm in $d=2$, while in $d=3$ we will revert to a simpler truncation on matrix element magnitudes alone.

\subsection{$d=2$}
\begin{figure}[ht]
    \centering
        \includegraphics[width=0.2\linewidth]{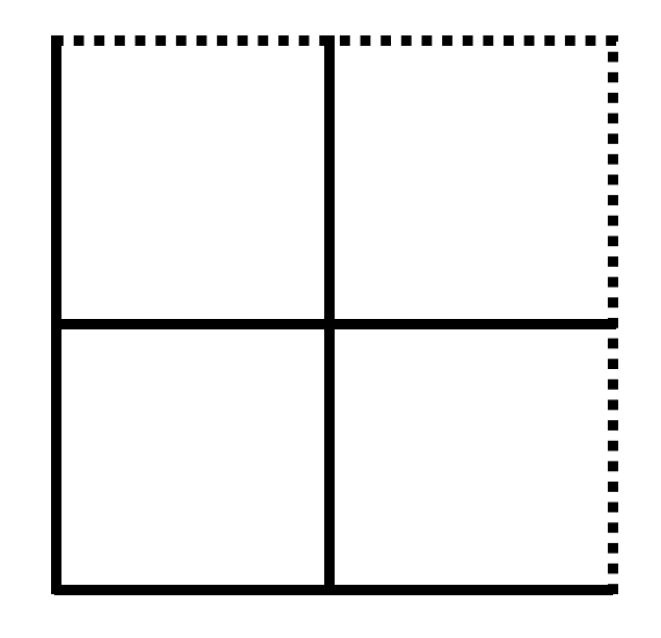}
    \caption{The $2\times 2$-plaquette lattice with PBCs.}
    \label{d2lat}
\end{figure}

We turn now to the simplest 2D lattice, $2\times 2$ plaquettes with periodic boundary conditions. 
The lattice is shown in  Fig.~\ref{d2lat}. 

\begin{figure}[ht]
    \centering
    \begin{subfigure}[t]{0.5\linewidth}
        \centering
        \includegraphics[width=\linewidth]{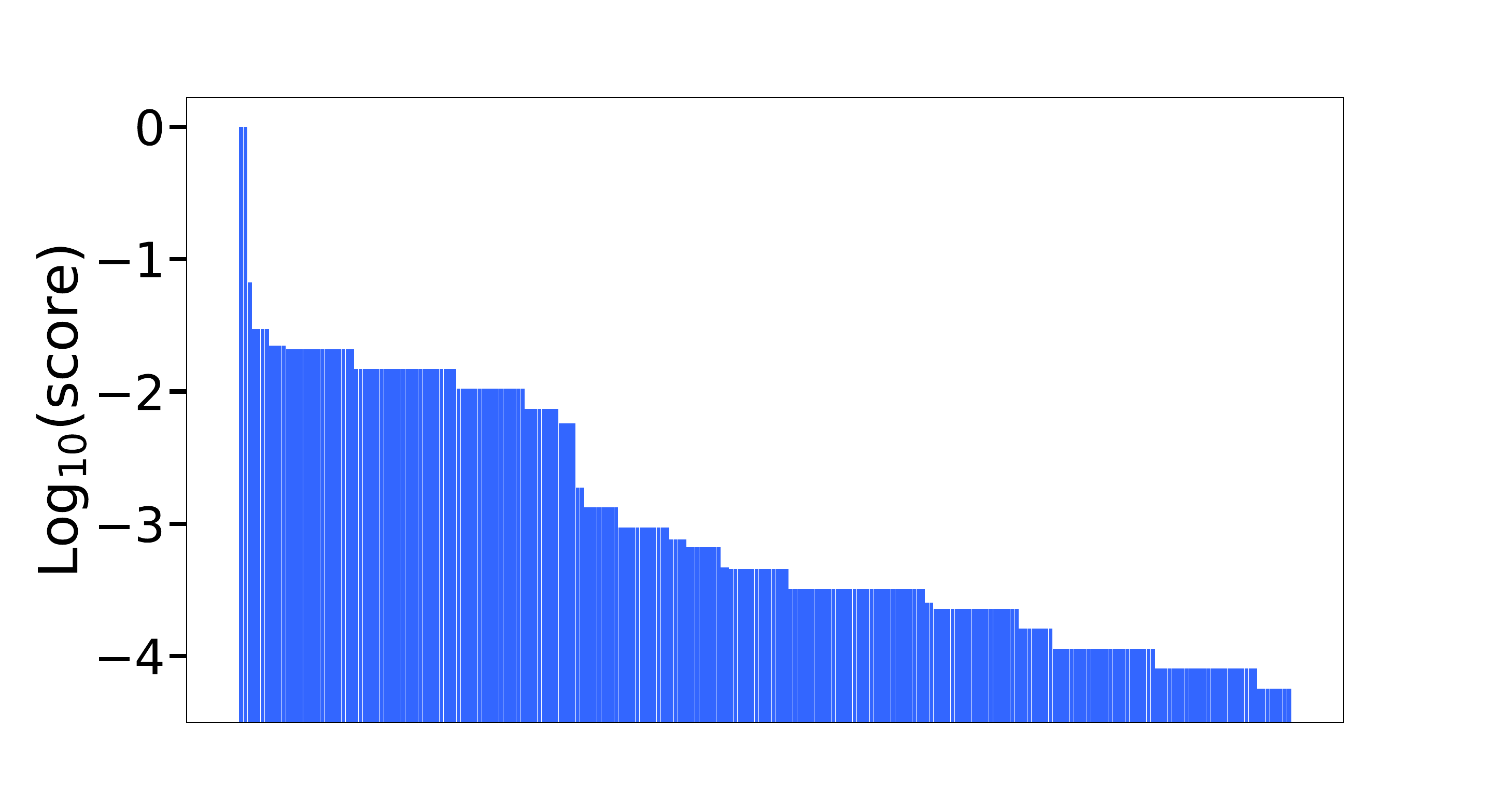}
    \end{subfigure}%
    ~
    \begin{subfigure}[t]{0.5\linewidth}
        \centering
        \includegraphics[width=\linewidth]{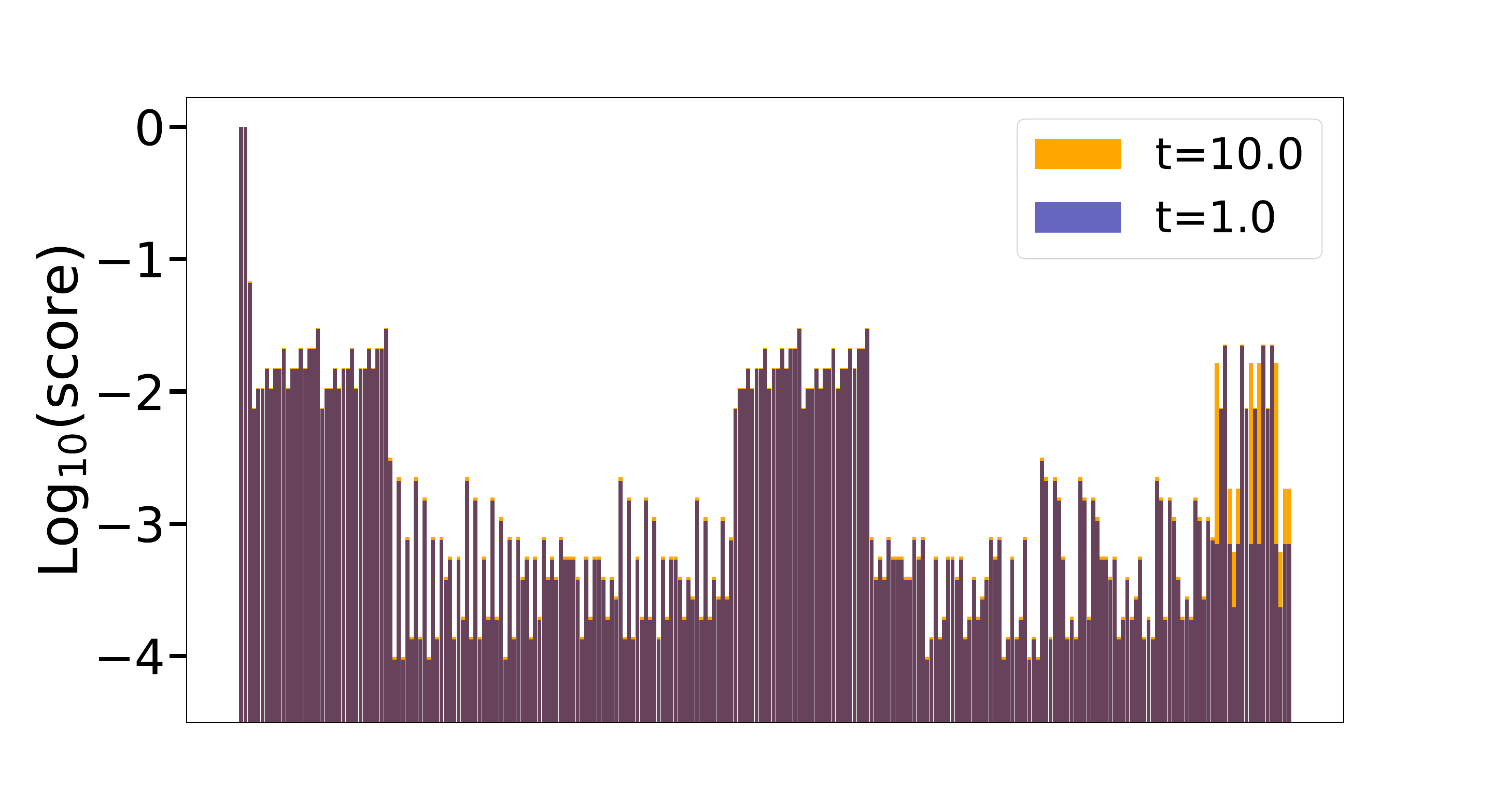}
    \end{subfigure}%
    \caption{Left: Matrix element scores for the $2\times 2$-plaquette lattice with $g=1$ and $T_1$ irrep truncation. Elements are ordered by descending score. Right: Matrix element scores for the same lattice and parameters, with two different time ranges ($T=1$, blue, and $T=10$, orange) used in the sampling. Here the matrix element ordering is arbitrary but fixed between the two values of $T$. Substantial overlap (purple) indicates that the scoring is largely insensitive to the sampling duration.}
    \label{figd2score}
\end{figure}

\begin{figure}[ht]
\centering
        \includegraphics[width=0.9\linewidth]{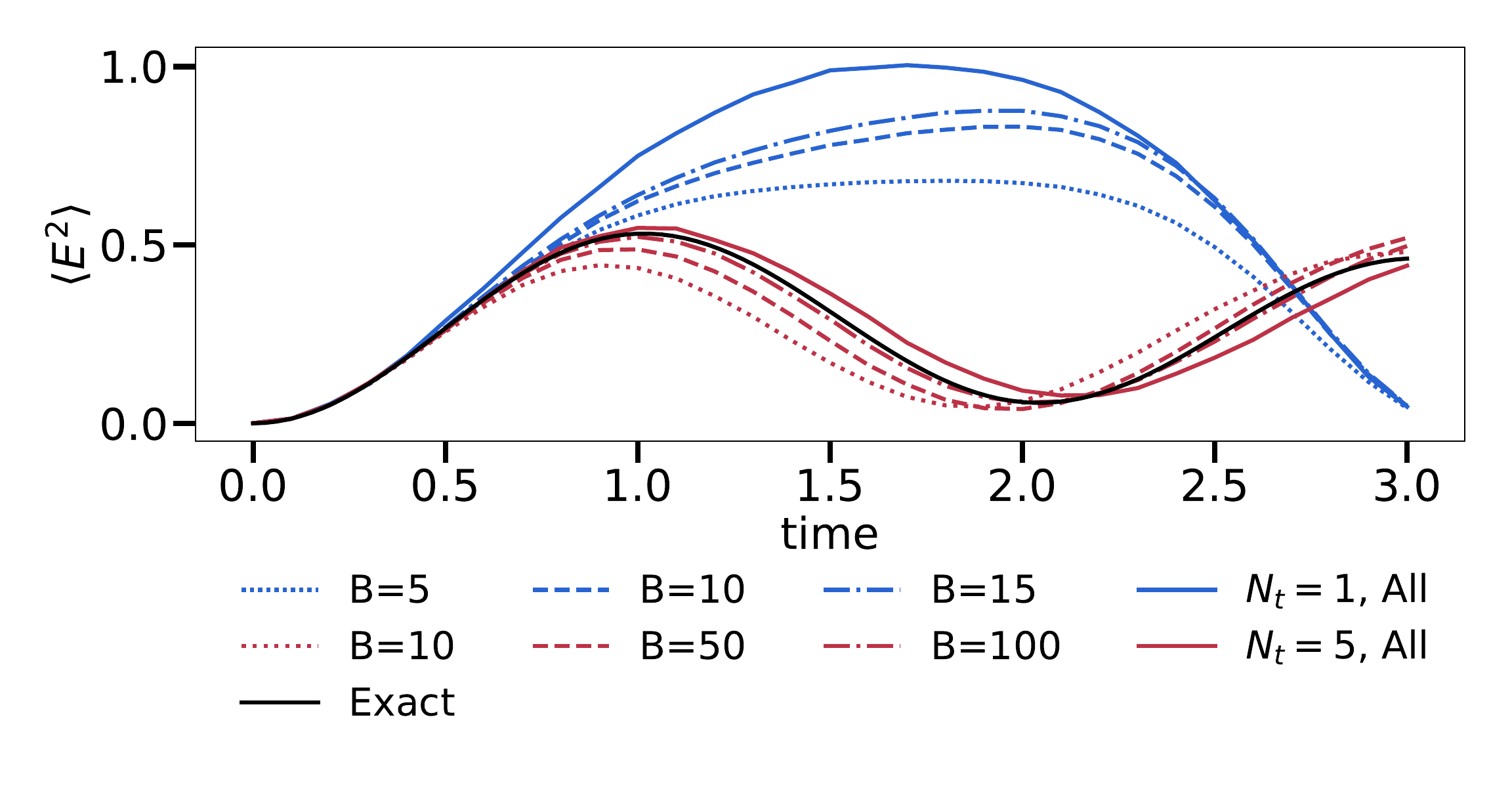}
    \caption{Time evolution of the link-averaged electric Casimir operator, starting from the electric vacuum on the $2\times 2$-plaquette lattice in the $T_1$ irrep truncation. Different curves correspond to different numbers of Trotter steps $N_t$ and truncations on the magnetic matrix element scores shown in Fig.~\ref{figd2score}; only the elements with the $B$ largest scores are kept in the truncation. }
    \label{figd2scoreevolution}
\end{figure}

We apply the plaquette matrix element scoring procedure described above for $d=3/2$. The left-hand panel of Fig.~\ref{figd2score} shows the distribution of scores for $T \! = \! 3$ again taken over 5 samples. Similar to $d=3/2$, out of the 835 matrix elements describing the evolution of $2 \times 2$-plaquette, only 247 describe transitions that can be reached from time evolution from the vacuum. Again we observe clear hierarchies among the matrix elements.  Fig.~\ref{figd2scoreevolution} 
compares the Trotter evolution of the electric vacuum in the $T_1$  truncations at $g=1$. Trotter error for $N_t=1$ becomes significant at times of order one in lattice units. For earlier times, a truncation to the transitions with the $B=5$ largest scores gives a good match to the exact evolution. For  $N_t=5$, reasonable agreement with the exact evolution is obtained when retaining on order of tens of the dominant transitions.

The scoring algorithm has a time duration parameter $T$. Scores are determined by sampling over a set of times up to $T$. The relative insensitivity of the scores to $T$ is illustrated in the right-hand panel of Fig.~\ref{figd2score}. We see that only a few of the scores evolve appreciably as the duration is increased by an order of magnitude.

\subsubsection{Noisy emulation}
\begin{figure}[ht]
    \centering
    \includegraphics[width=0.75\linewidth]{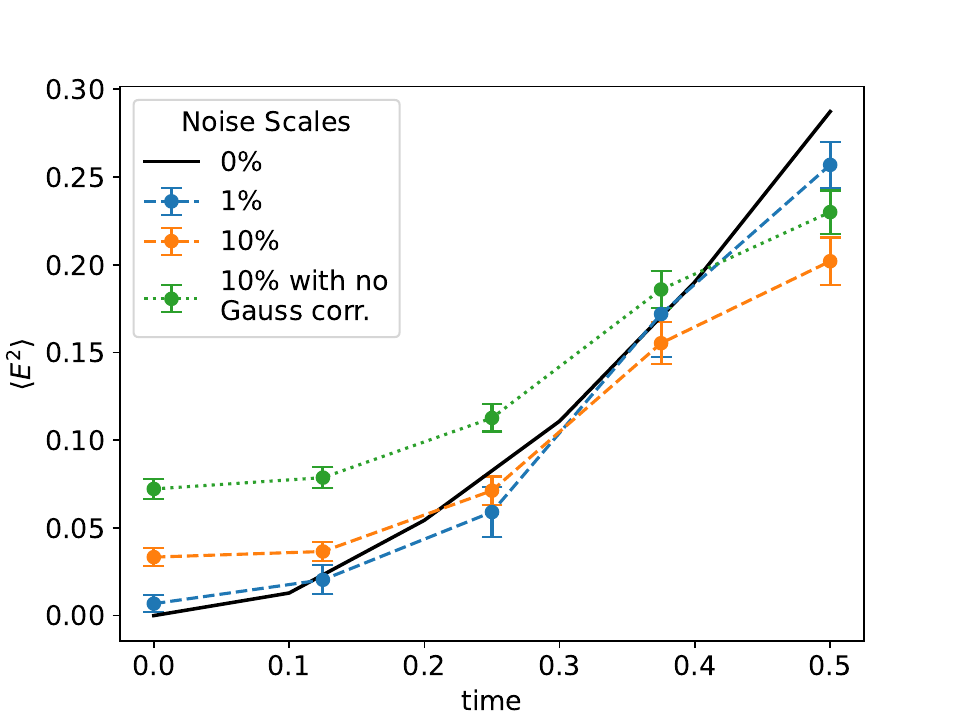}
    \caption{Quantinuum H2 emulator results for the evolution of the electric vacuum, $g=1$, and one Trotter step on the $2\times 2$ lattice with PBCs and the $T_1$ irrep truncation. Ancillas are used in the decomposition of multi-controlled gates to reduce two-qubit gate counts. We apply a rudimentary form of error mitigation, vetoing shots that correspond to unphysical states. Dashed curves correspond to different scale factors applied to the noise. Error bars on these results reflect shot noise. (200 shots were used for the first four ``1\%" points, increased to 1000 shots for the final point to reduce noise. 1000 shots were used for all ``10\%" points.) The black curve corresponds to the noiseless simulation. }
    \label{fig:noisy}
\end{figure}

The circuits developed in this work are best suited for quantum platforms with substantial qubit connectivity: although the Hamiltonian is local in the sense of coupling nearest-neighbor links and plaquettes, the plaquette Hilbert space can span tens of qubits, which are all coupled together. In Fig.~\ref{fig:noisy}  we report results of noisy simulations of the $2\times 2$ PBC lattice of Fig.~\ref{d2lat}, with T1 irrep truncation and $B=3$ transition truncation, using the Quantinuum System Model H2 emulator~\cite{Quantinuum}. H2 is a 56 qubit trapped ion device with all-to-all connectivity. The emulator allows scaling the noise levels and we present results with all noise sources uniformly dialed to $1\%$ and $10\%$. As a mild form of error mitigation, we veto shots that go outside the physical link Hilbert space (with irrep truncation $T_1$, the two-qubit $11$ bitstring is not assigned an irrep), and then to veto shots that fail the Gauss law at some vertex.  At $1\%$ the noise is insignificant for this circuit. Above $10s$ of percent, the noise is not effectively mitigated by the simple mitigation strategy.

\subsection{$d=3$}

\begin{figure}[ht!]
    \centering
        \includegraphics[width=0.2\linewidth]{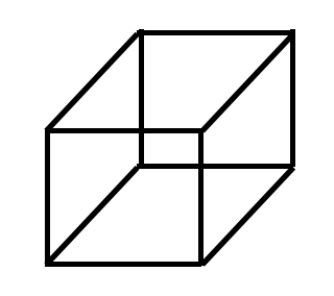}
    \caption{The cube lattice with OBCs.}
    \label{cubeobc}
\end{figure}

Although we are able to build circuits for larger lattices in $d=3$, our ability to execute noiseless statevector simulations is limited to $n_{\text{qubit}}\lesssim 40$. A simple lattice where we can perform simulations is the single cube with open boundary conditions, shown in Fig.~\ref{cubeobc}. 

\begin{figure}[ht!]
    \centering
    \begin{subfigure}[ht]{0.6\linewidth}
        \centering
        \includegraphics[width=\linewidth]{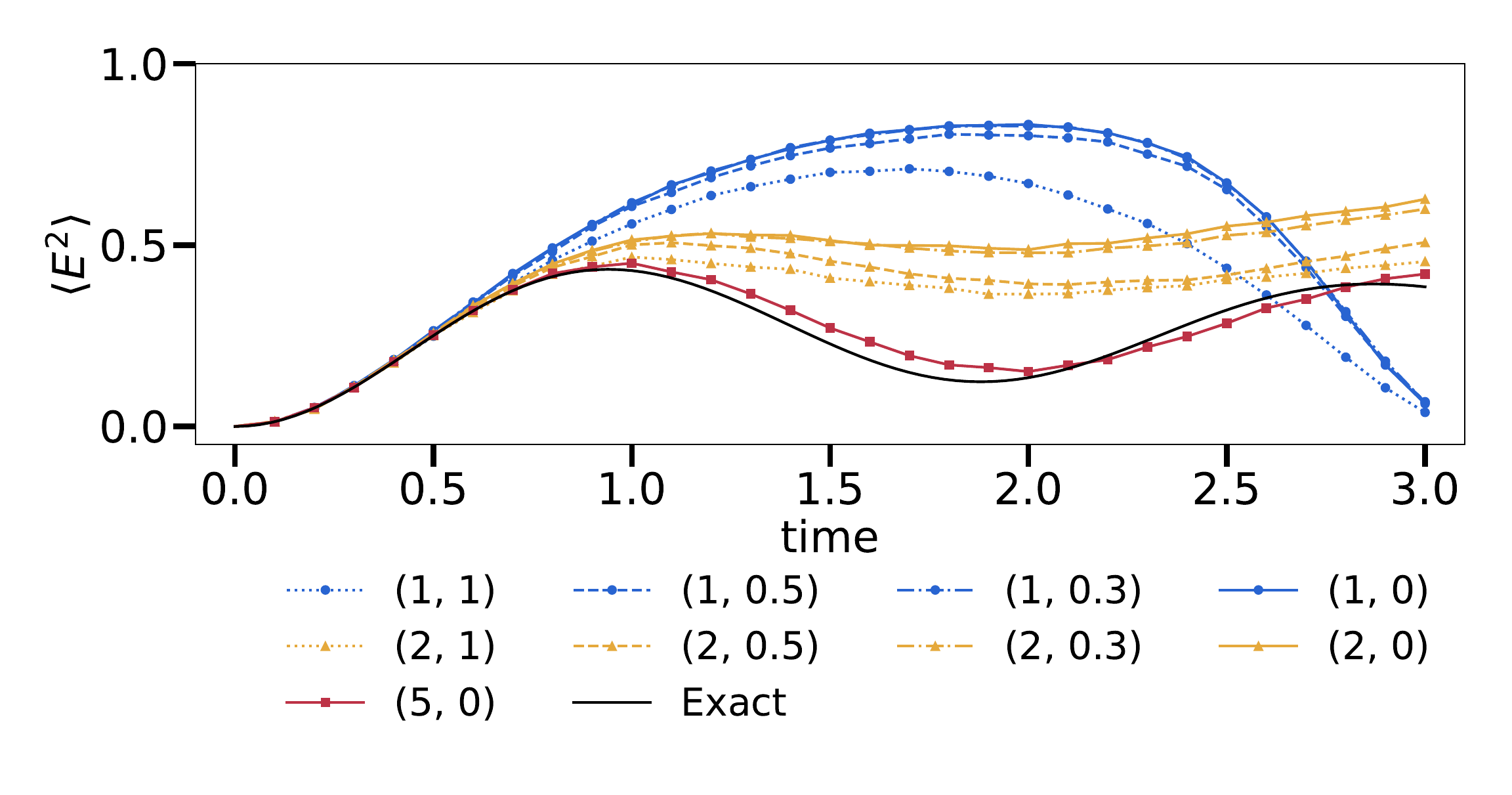}
    \end{subfigure}
    \\
    \begin{subfigure}[ht]{0.6\linewidth}
        \centering
        \includegraphics[width=\linewidth]{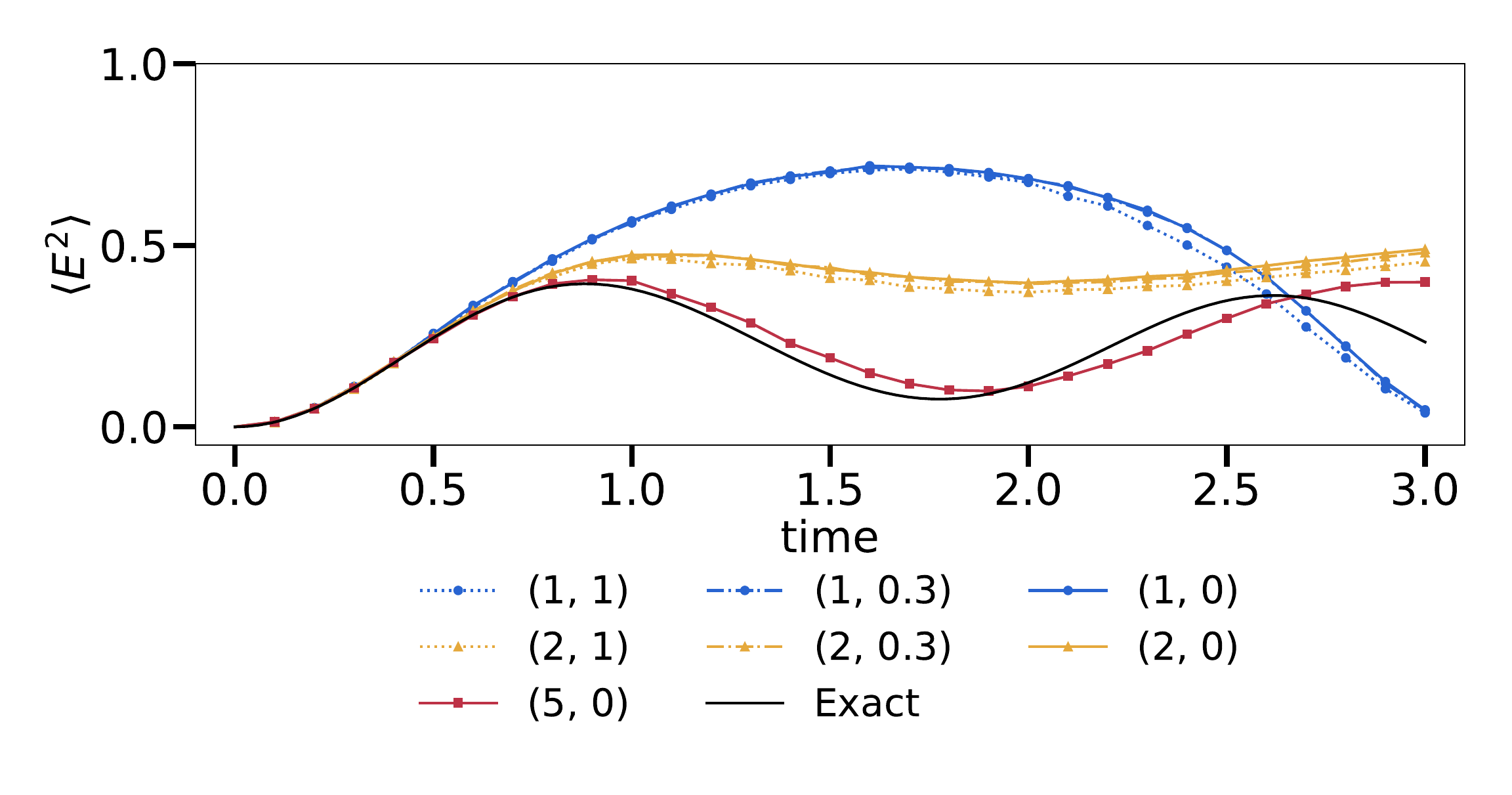}
    \end{subfigure}
    \caption{Link-averaged electric Casimir for the cube lattice with OBC in $T_1$ (top) and $T_1'$ (bottom). Different curves correspond to varying the number of Trotter steps and the cutoff on magnetic matrix elements ($N_{\text{t}}$, cutoff). Top: A cutoff of 0 means all 81 magnetic matrix elements were used. Cutoffs of $(0.3,0.5,1)$ correspond to $(71,27,3)$ matrix elements, respectively. Bottom: A cutoff of 0 means all 17 magnetic matrix elements were used. Cutoffs of $(0.3,1)$ correspond to $(15,3)$ matrix elements, respectively.}
    \label{fig:cubeobc}
\end{figure}

In this case, we opt for a simpler truncation of magnetic transitions, cutting on the magnitude of matrix elements alone, rather than the score defined in Eq.~(\ref{eq:score}). Results for $g=1$, $T_1$ truncation are shown in the top panel of Fig.~\ref{fig:cubeobc}. We also examine the $T_1'$ singlet truncation in the bottom panel of Fig.~\ref{fig:cubeobc}.  
We find that for this lattice the full $T_1'$ simulation (17 transitions) well-approximates the full $T_1$ truncation (81 transitions).

\subsubsection{Tensor networks}
\begin{figure}[ht]
\centering
\includegraphics[width=0.49\linewidth]{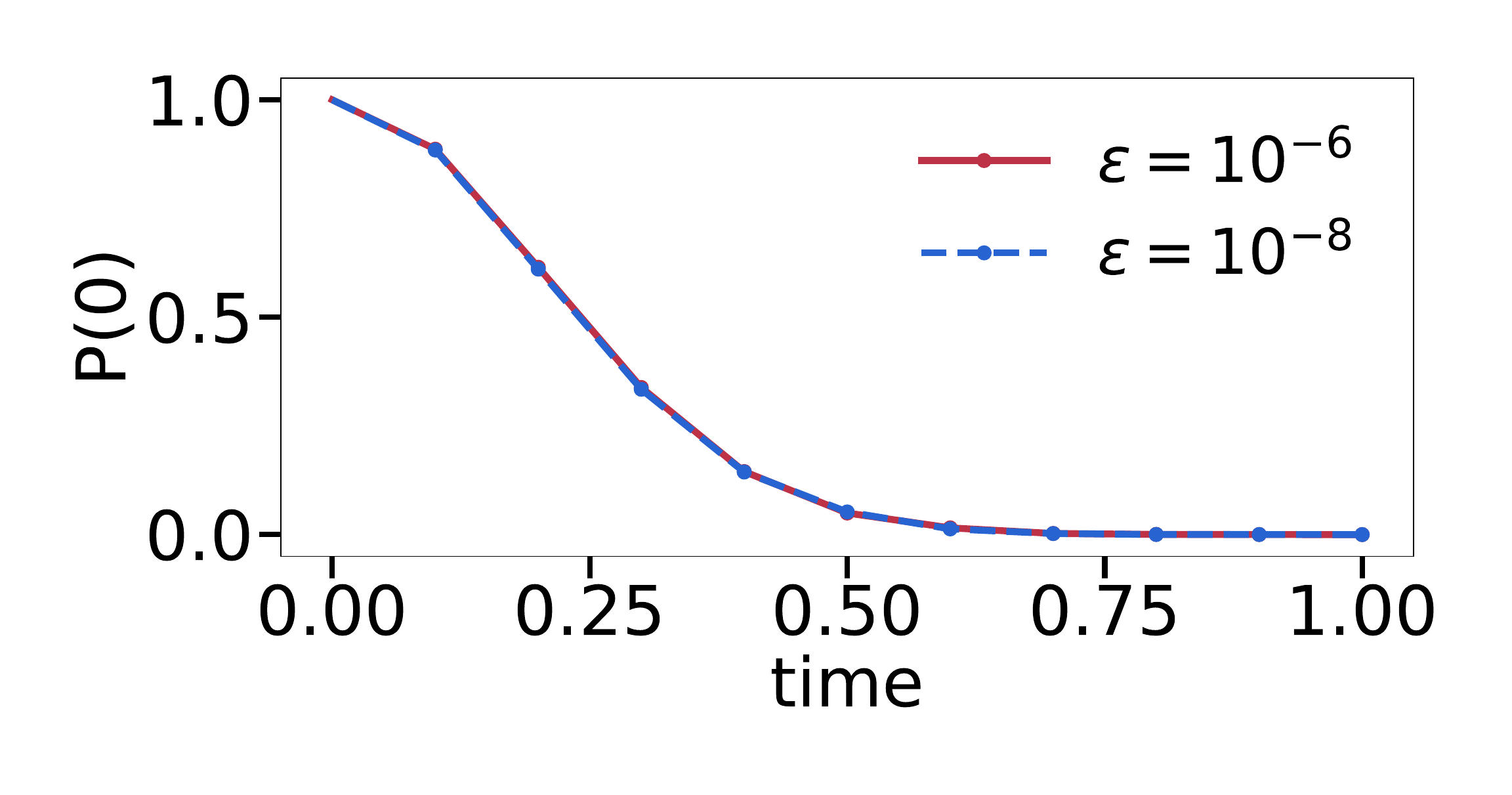}
\includegraphics[width=0.49\linewidth]{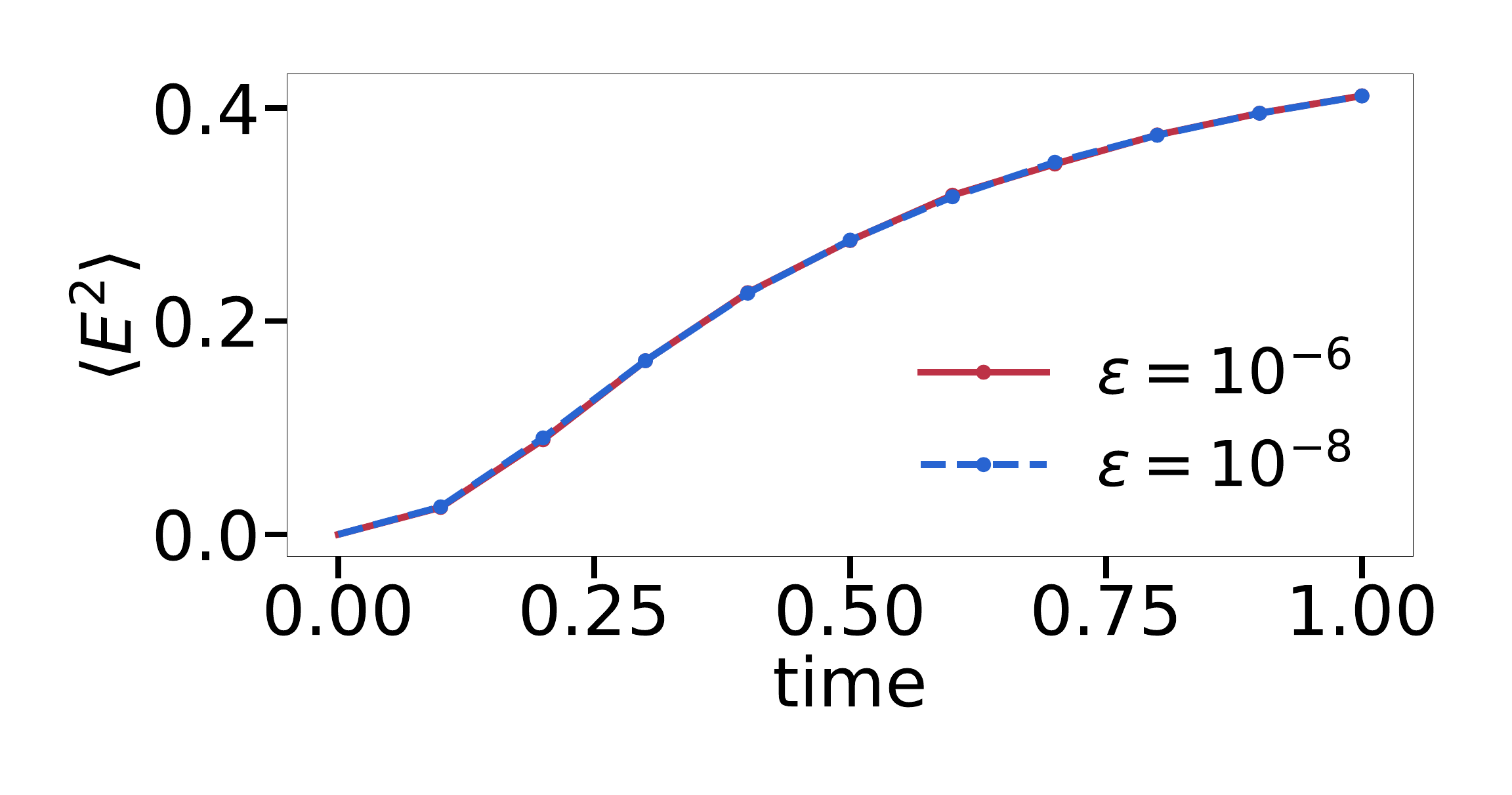}
\caption{Vacuum-persistence probability (left) and link-averaged electric Casimir (right) for the $2 \times 2 \times 2$ (24-plaquette) lattice with PBCs and $T'_1$ truncation, computed using matrix product state simulation, one Trotter step, with truncation threshold set to $\epsilon=10^{-6}$ (red) and $10^{-8}$ (blue, dashed). The curves are essentially coincident. }
\label{fig:2cubed_mps}
\end{figure}

Higher $n_{\text{qubit}}$ can be accessed in classical simulations using tensor network approximations. We perform preliminary analysis of a $2\times2\times2$ (24 plaquette) lattice with PBCs and a maximum-element truncation of the magnetic transitions, using the out-of-the-box matrix product state (MPS) method included in Qiskit. This is a 48 qubit lattice, out of range of statevector simulation. Fig.~\ref{fig:2cubed_mps} shows the electric Casimir at $g=1$ for a single Trotter step and TN thresholds of $10^{-6}$ and $10^{-8}$. We observed convergence for threshold $\lesssim 10^{-6}$. 

The 48 qubit implementation of the circuit does not use ancillas in the MCU decomposition. Consequently it is a deep circuit, well out of range of current NISQ hardware. With ancillas, the depth is reduced, but the qubit count (66) exceeds that of current devices with all-to-all connectivity. It represents an interesting target for next-generation hardware. This lattice at higher $N_t$, and the $3\times3\times 3$ lattice, may be useful benchmarks for comparing classical TN methods to quantum simulations.  Further tensor treatments of the circuits constructed here will be presented elsewhere.

\section{Future directions}
\label{future}
We close with a brief summary of directions for future development. 

On the physics front, it is necessary to incorporate quarks and a $\theta$ term; to develop ground state preparation methods;  to implement measurement of expectation values of plaquette operators; and to study the continuum limit. For example, unlike the pure gauge analysis here, inclusion of quarks will require allowing gauge links at sites to fuse into nontrivial irreps. Matrix elements of localized operators in the Hamiltonian can still be classically pre-computed using generalized Clebsch-Gordan coefficients. Similarly, minimizing the resource requirements to approach the continuum limit will require the addition of improvement terms~\cite{Lepage:1992xa} (for a quantum simulation study, see~\cite{Carena:2022kpg}.) In the approach taken here, incorporating at least classical improvement terms amounts to computing master formulas for the matrix elements of some additional operators. 

Minimizing the quantum resource requirements for physically relevant simulations will require further advances in circuit compression and development of new approximation schemes. 
We have begun preliminary explorations of some algorithms; for example, binning magnetic rotations by magnitude and rebasing the states so that the rotations in a bin can be fused into a smaller set of multi-controlled unitaries. This will be reported on elsewhere. We have also described here some crude methods for assessing ``which matrix elements really matter" in a given simulation. In our examples the answer is generally ``much less than all of them," but clearly there is much more to be understood about this structure and how to determine it a priori.

There are several natural directions for extending the functionality of our modular circuit generation code. These include extending circuit-generation capability from $SU(3)$ to $SU(N)$ utilizing the methods outlined in Sec.~\ref{cgsec}, integrating the pre-computation of matrix elements into the simulation pipeline to support dynamic circuit generation in arbitrary lattice dimensionalities and  truncations, and the implementation of higher-order Trotterization in simulation circuits.

Finally, as is well-known but abundantly clear in the circuits we have constructed,  qubit platforms with one- and two-qubit gates furnish sufficient but quite unnatural architectures for simulating lattice gauge theories. Qudit systems, or qubits with native multi-controlled gates, could offer significant reductions in circuit depth for these kinds of simulation problems. It would be interesting to sharpen the precise hardware advances that would be most impactful.

\section*{Acknowledgments}
We thank Aida El-Khadra for helpful conversations and collaboration in the early stages of this work.

This work was supported by the U.S. Department of Energy,
Office of Science, Office of High Energy Physics Quantum Information Science Enabled Discovery (QuantISED) program. We gratefully acknowledge BlueQubit~\cite{BlueQubit} for enabling classical statevector simulations presented in this work. We also gratefully acknowledge the use of Quantinuum H1-1 and H2-1 emulators~\cite{Quantinuum}.

This work used the Delta~\cite{Delta} system at the National Center for Supercomputing Applications through allocation PHY230137 from the Advanced Cyberinfrastructure Coordination Ecosystem: Services \& Support (ACCESS) program, which is supported by National Science Foundation grants \#2138259, \#2138286, \#2138307, \#2137603, and \#2138296.
The Delta advanced computing and data resource is supported by the National Science Foundation (award OAC 2005572) and the State of Illinois. Delta is a joint effort of the University of Illinois Urbana-Champaign and its National Center for Supercomputing Applications.

\bibliographystyle{utphys}
\bibliography{refs}
~\\

\appendix
\section{Master Formula}

\begin{align}
  \hspace{-0.5in}& \bra{\omega_n} \Box\{ P(\vec{s},\vec{e}_i,\vec{e}_j) \} \ket{\omega_m}  = \nonumber\\
  &~~ \sqrt{\frac{\dim(R_{\ell_0}^m) \dim(R_{\ell_1}^m) \dim(R_{\ell_2}^m) \dim(R_{\ell_3}^m)}{\dim(R_{\ell_0}^n) \dim(R_{\ell_1}^n) \dim(R_{\ell_2}^n) \dim(R_{\ell_3}^n)}} \prod_{s \notin \mathcal{S}(P)} \delta_{\Gamma_s^n, \Gamma_s^m} \prod_{\ell \notin \mathcal{L}(P)} \delta_{R^n_\ell, R^m_\ell} \nonumber\\ 
  &~~  \sum_{\sigma_1=1}^{\dim(f)} \sum_{\lambda_{\ell_1^-}^m=1}^{\dim(R_{\ell_1}^m)} \sum_{\lambda_{\ell_1^-}^n=1}^{\dim(R_{\ell_1}^n)} \sum_{\lambda_{\ell_4^-}^m=1}^{\dim(R_{\ell_4}^m)} \sum_{\lambda_{\ell_4^-}^n=1}^{\dim(R_{\ell_4}^n)} \sum_{\vec{\chi} \in \vec{C}_{s_1}} \phi(\sigma_1) \phi(\lambda_{\ell_1^-}^m) \phi(\lambda_{\ell_1^-}^n) \phi(\lambda_{\ell_4^-}^m) \phi(\lambda_{\ell_4^-}^n)\nonumber \\
  &~~ \times  \BraKet{R_{\ell_1}^n,\lambda_{\ell_1^-}^n}{(R_{\ell_1}^m,\lambda_{\ell_1^-}^m) \otimes (f,\sigma_1)} \ \BraKet{R_{\ell_4}^n,\lambda_{\ell_4^-}^n}{(R_{\ell_4}^m,\lambda_{\ell_4^-}^m) \otimes (\bar{f},\tilde{\sigma}_1)}\nonumber \\
  &~~ \times  \BraKet{1,\Gamma_{s_1}^m}{(\bar{R}_{\ell_1}^m,\tilde{\lambda}_{\ell_1^-}^m) \otimes (\bar{R}_{\ell_4}^m,\tilde{\lambda}_{\ell_4^-}^m) \otimes (\vec{C}_{s_1},\vec{\chi})} \ \BraKet{1,\Gamma_{s_1}^n}{(\bar{R}_{\ell_1}^n,\tilde{\lambda}_{\ell_1^-}^n) \otimes (\bar{R}_{\ell_4}^n,\tilde{\lambda}_{\ell_4^-}^n) \otimes (\vec{C}_{s_1},\vec{\chi})}\nonumber \\ 
  &~~  \sum_{\sigma_2=1}^{\dim(f)} \sum_{\lambda_{\ell_1^+}^m=1}^{\dim(R_{\ell_1}^m)} \sum_{\lambda_{\ell_1^+}^n=1}^{\dim(R_{\ell_1}^n)} \sum_{\lambda_{\ell_2^-}^m=1}^{\dim(R_{\ell_2}^m)} \sum_{\lambda_{\ell_2^-}^n=1}^{\dim(R_{\ell_2}^n)} \sum_{\vec{\chi} \in \vec{C}_{s_2}} \phi(\lambda_{\ell_2^-}^m) \phi(\lambda_{\ell_2^-}^n)\nonumber \\
  &~~ \times  \BraKet{R_{\ell_1}^n,\lambda_{\ell_1^+}^n}{(R_{\ell_1}^m,\lambda_{\ell_1^+}^m) \otimes (f,\sigma_2)} \ \BraKet{R_{\ell_2}^n,\lambda_{\ell_2^-}^n}{(R_{\ell_2}^m,\lambda_{\ell_2^-}^m) \otimes (f,\sigma_2)} \nonumber\\
  &~~ \times  \BraKet{1,\Gamma_{s_2}^m}{(R_{\ell_1}^m,\lambda_{\ell_1^+}^m) \otimes (\bar{R}_{\ell_2}^m,\tilde{\lambda}_{\ell_2^-}^m) \otimes (\vec{C}_{s_2},\vec{\chi})} \ \BraKet{1,\Gamma_{s_2}^n}{(R_{\ell_1}^n,\lambda_{\ell_1^+}^n) \otimes (\bar{R}_{\ell_2}^n,\tilde{\lambda}_{\ell_2^-}^n) \otimes (\vec{C}_{s_2},\vec{\chi})}\nonumber \\
  &~~  \sum_{\sigma_f=1}^{\dim(f)} \sum_{\lambda_{\ell_3^+}^m=1}^{\dim(R_{\ell_3}^m)} \sum_{\lambda_{\ell_3^+}^n=1}^{\dim(R_{\ell_3}^n)} \sum_{\lambda_{\ell_2^+}^m=1}^{\dim(R_{\ell_2}^m)} \sum_{\lambda_{\ell_2^+}^n=1}^{\dim(R_{\ell_2}^n)} \sum_{\vec{\chi} \in \vec{C}_{s_3}} \phi(\sigma_3) \nonumber\\
  &~~ \times  \BraKet{R_{\ell_3}^n,\lambda_{\ell_3^+}^n}{(R_{\ell_3}^m,\lambda_{\ell_3^+}^m) \otimes (\bar{f},\tilde{\sigma}_3)} \ \BraKet{R_{\ell_2}^n,\lambda_{\ell_2^+}^n}{(R_{\ell_2}^m,\lambda_{\ell_2^+}^m) \otimes (f,\sigma_3)} \nonumber\\
  &~~ \times  \BraKet{1,\Gamma_{s_3}^m}{(R_{\ell_3}^m,\lambda_{\ell_3^+}^m) \otimes (R_{\ell_2}^m,\lambda_{\ell_2^+}^m) \otimes (\vec{C}_{s_3},\vec{\chi})} \ \BraKet{1,\Gamma_{s_3}^n}{(R_{\ell_3}^n,\lambda_{\ell_3^+}^n) \otimes (R_{\ell_2}^n,\lambda_{\ell_2^+}^n) \otimes (\vec{C}_{s_3},\vec{\chi})} \nonumber\\ 
  &~~  \sum_{\sigma_4=1}^{\dim(f)} \sum_{\lambda_{\ell_3^-}^m=1}^{\dim(R_{\ell_3}^m)} \sum_{\lambda_{\ell_3^-}^n=1}^{\dim(R_{\ell_3}^n)} \sum_{\lambda_{\ell_4^+}^m=1}^{\dim(R_{\ell_4}^m)} \sum_{\lambda_{\ell_4^+}^n=1}^{\dim(R_{\ell_4}^n)} \sum_{\vec{\chi} \in \vec{C}_{s_4}} \phi(\lambda_{\ell_3^-}^m) \phi(\lambda_{\ell_3^-}^n) \nonumber\\
  &~~ \times  \BraKet{R_{\ell_3}^n,\lambda_{\ell_3^-}^n}{(R_{\ell_3}^m,\lambda_{\ell_3^-}^m) \otimes (\bar{f},\tilde{\sigma}_4)} \ \BraKet{R_{\ell_4}^n,\lambda_{\ell_4^+}^n}{(R_{\ell_4}^m,\lambda_{\ell_4^+}^m) \otimes (\bar{f},\tilde{\sigma}_4)} \nonumber\\
  &~~ \times  \BraKet{1,\Gamma_{s_4}^m}{(\bar{R}_{\ell_3}^m,\tilde{\lambda}_{\ell_3^-}^m) \otimes (R_{\ell_4}^m,\lambda_{\ell_4^+}^m) \otimes (\vec{C}_{s_4},\vec{\chi})} \ \BraKet{1,\Gamma_{s_4}^n}{(\bar{R}_{\ell_3}^n,\tilde{\lambda}_{\ell_3^-}^n) \otimes (R_{\ell_4}^n,\lambda_{\ell_4^+}^n) \otimes (\vec{C}_{s_4},\vec{\chi})}.
  \label{masterformulafull}
\end{align}

\end{document}